\documentclass[12pt]{article}
\pdfoutput=1
\usepackage{jheppub}

\usepackage{amsmath} 
\usepackage{amssymb,amsfonts}
\usepackage[dvipsnames]{xcolor}
\usepackage{graphicx}
\usepackage[utf8]{inputenc}

\usepackage{subcaption}  
\usepackage{graphicx}
\usepackage[utf8]{inputenc}
\usepackage{amssymb}
\usepackage{verbatim}
\usepackage{amsmath}
\usepackage{color}

\usepackage{xcolor}

\newcommand{\be}{\begin{equation}}
\newcommand{\ee}{\end{equation}}
\newcommand{\bea}{\begin{eqnarray}}
\newcommand{\eea}{\end{eqnarray}}

\newcommand{\eps}{\epsilon}

\newcommand{\morder}[1]{{\cal O}\left(#1 \right)}
\newcommand{\nb}{\mathrm{NB}}
\newcommand{\nh}{\mathrm{NH}}

\definecolor{oblue} {RGB}{91,125,191}

\title{An analytic approach for holographic entanglement entropy at (quantum) criticality}

\affiliation[a]{  Department of Physics, Fergusson College (Autonomous), Savitribai Phule Pune University, Pune - 411004, India.}
\affiliation[b]{Asia Pacific Center for Theoretical Physics,   Pohang 37673,   Korea }
\affiliation[c]{Department of Physics, Pohang University of Science and Technology,   Pohang 37673,   Korea }
\affiliation[d]{Institute of Theoretical Physics, Chinese Academy of Sciences, Beijing 100190, China}

\author[a]{Parul Jain}
\author[b,c,d]{ Matti J\"arvinen}
\date{July 2025}

\emailAdd{parul.jain@fergusson.edu}
\emailAdd{mattijarvinen@itp.ac.cn}

\preprint{APCTP Pre2025 - 023}

\abstract{We consider holographic entanglement entropy in AdS black hole backgrounds by using the limit of large number of dimensions. By dividing the geometry to two patches (with one patch covering the vicinity of the black hole horizon and another covering the other regions), we are able to obtain fully analytic expressions for the entropy when the entanglement region is a strip. We argue that apart from conformal field theories at finite temperature in high number of dimensions, our method works for nearly critical theories in 3+1 dimensions. In the case of extremal black holes, dual to quantum critical systems, the results take a particularly simple form. We also comment on the case of soliton geometries. Finally, we analyze entanglement entropy for wide strips, and propose a general formula for the first subleading term in the expansion of the entropy in (inverse) system size for generic entanglement regions. }

\begin{document}
\maketitle
%\tableofcontents
\newpage

\section{Introduction}

Black hole analysis is a broad and an important field in theoretical physics, incorporating various approaches to studying black holes using mathematical, numerical, and observational methods. Different areas of research address different aspects of black hole physics, from their formation and dynamics to their observational signatures and quantum properties. Holography, which is a duality between field theory and gravity, is one of the most widely used backgrounds to study diverse features of black holes such as perturbations, instabilities, gravitational waves, mergers, quantum gravity, black hole thermodynamics etc. One of the pertinent research direction which uses holographic tools is to explore the relation between quantum information and black holes, e.g., connection of  different entanglement measures such as entanglement entropy, negativity, purification etc. with Bekenstein-Hawking entropy, Hawking radiation, phase transition and critical phenomena in black holes. An important parameter in all these analyses, apart from the black hole mass, charge and rotation is the geometry of the black hole.
%%%%%%%%%%%%%%%%%%%%%%%%%%%%%%%%%%%%%

A method which often greatly simplifies~\cite{Emparan:2020inr} analysis of black hole geometries is to study the expansion in $1/D$, where $D$ is the total number of dimensions in the gravity theory being considered. In the limit of large $D$, the black hole becomes membrane-like: the space far away from the horizon looks essentially flat, and the effect of the horizon to the geometry is localized in the region within the distance $\sim 1/D$ from the horizon. This also means that the regimes far away from the horizon and near the horizon become essentially decoupled and can be analyzed separately, both for the geometry and its fluctuations. Actually, as it turns out, the two description of the geometry (the flat geometry far from the horizon and the geometry within the distance $\sim 1/D$ from the horizon) have overlapping range of validity. This may lead to enhanced analytic control: for example, one can often solve analytically the fluctuations of the black hole in both regions, and match the solutions in the overlapping region, which gives rise to analytic control of the fluctuations for the whole black hole at large $D$.

One of the main purposes of this article is to show that these ideas also apply to the computation of the holographic entanglement entropy in the limit of large $D$.
Entanglement entropy is a pure state entanglement measure and is given by the von Neumann entropy 
\begin{equation}
 S_A = -\mathrm{Tr}_A\rho_A\log\,\rho_A\ . \label{vne}
\end{equation}
In case of field theories, one employs the replica method to calculate the entanglement entropy whereas for holographic scenarios, we use the 
Ryu--Takayanagi formalism \cite{Ryu:2006ef,Ryu:2006bv}. This formalism says that the entanglement entropy for a given region $A$ of the boundary field theory which is bounded by $\partial A$ is dictated by the area of the minimal surface $\Gamma_{A}$ which is anchored to $\partial A$
\begin{equation}
 S_A = \frac{\mathcal{A}_{\Gamma_A}}{4G_N^{(d+1)}}\ \ , \ \   \mathcal{A}_{\Gamma_A} = \int_{\Gamma_A} d^{d-1}x\sqrt{\det g_{d-1}} \ , \label{eq:hee2}
\end{equation}
where $g_{d-1}$ is the induced metric on $\Gamma_A$, which is 
{required to be minimal}
and it's boundary is homologous to the boundary of $A$.

In the simplest application, one considers  AdS$_D$ Schwarzschild black hole   geometries, dual to CFT$_d$ with $D=d+1$ at finite temperature. We focus on AdS spaces in the Poincare patch, so that our black holes extend to all values of the spacetime coordinates, and the dual CFT lives on a flat Euclidean or Minkowski metric. In this context we call the regime of the geometry far from the horizon the near boundary (NB) region, as being far from the horizon also means being close to the boundary of the AdS geometry. The complementing region is then called the near horizon (NH) region. As it turns out, one can indeed carry out the same procedure for the holographic entanglement entropy as what is standard for fluctuations of large-$D$ lack holes: one can find analytic solutions for entanglement surfaces in various black hole geometries  both in the NB and NH regions, and match them in the middle in the regime of overlapping validity. The range of overlap is enhanced at large $D$,  giving a full analytic control on the results in this limit.

Our approach complements earlier analyses of entanglement entropy at large $D$. In particular,~\cite{Garcia-Garcia:2015emb,Giataganas:2021jbj} addressed (among other observables) entanglement entropy for strips in CFTs in the limit of large $D$ by employing a perturbative expansion in the finite-temperature corrections. As we shall see, these results form one part of the method proposed here, i.e., the results in the NB region but will be extended by the analysis of the NH region. 
Moreover, the large $D$ limit was used to study mutual information in global AdS backgrounds in~\cite{Colin-Ellerin:2019vst}. 

Apart from works on holographic quantum information, large $D$ methods have been applied to study other observables in the context of holography. Considering dynamical black hole backgrounds at large $D$ quite in general leads to membrane-like picture for the black hole, where the dynamics is described through a hydrodynamic theory of this membrane, and also has an holographic interpretation~\cite{Emparan:2015hwa,Emparan:2015gva,Bhattacharyya:2015dva,Dandekar:2016jrp,Andrade:2018zeb,Jarvinen:2024vdi}.  Other works include, in particular, different approaches to time-evolution~\cite{Casalderrey-Solana:2018uag}, momentum relaxation~\cite{Andrade:2015hpa}, turbulence~\cite{Rozali:2017bll}, and physics non-equilibrium steady states~\cite{Herzog:2016hob} (see also~\cite{Licht:2022rke}).

Interestingly, while we consider contributions to the entanglement entropy both from the NB and NH regions, one of our main results can be obtained without considering this division to two regions. Namely, we analyze the entanglement entropy $S_E$ for a strip of width $L$ in the limit of large width, $L \to \infty$ (or equivalently $S_E$ at fixed $L$ but in the limit of high temperature). 
The leading term in this expansions matches the thermal entropy of the entangling region. However, also the subleading term can be computed analytically if one also takes $D$ to be large~\cite{Giataganas:2021jbj}. This subleading term is interesting because it is known to be monotonic under renormalization group flows in Lorentz-covariant settings~\cite{Casini:2012ei,Casini:2016udt}, a property known as the ``area theorem''.  
As it turns out, the term solely arises from the NB geometry, so that the division into NB and NH regimes and matching is not required to obtain it. Actually, in this article, we argue that thanks to the term being NB dominated it can be computed analytically for quite general entangling regions in the limit of their large size, and the result is related to the free energy in these regions. For a general region $A$ in a CFT$_d$, our result for the leading and subleading terms is given by (see Sec.~\ref{sec:expansion}) 
\be \label{eq:SElargeAintro}
 S_E(A) \approx \mathrm{Vol}(A) s + \mathrm{Vol}(\partial A) \frac{2\pi}{d} \left( sT+\mu Q\right)
\ee
with corrections suppressed by the inverse of the region size and $1/d$. Here $s$ is the entropy density, $T$ is the temperature, $Q$ is a the charge of the black hole, and $\mu$ is the corresponding chemical potential.

One might think that applying a $1/D$ expansion in the number of dimensions takes one very far from physically interesting, e.g. 3+1 dimensional field theories, and interacting CFTs with $d>6$ are not even known to exist. However, this might not be the case for two reasons. First, the $1/D$ expansion usually converges relatively well, and for $D \ge 4$ one does not expect qualitative changes (such as phase transitions), so that the large $D$ limit may be used describe gravity at low $D$ in a reasonable approximation. Usually observables such as fluctuations of black holes can be well described down to rather low values such as $D=6$ by using expansions in $1/D$~\cite{Emparan:2015rva,Andrade:2018nsz}. The same applies to non-local observables (such as Wilson loops and entanglement entropy) in holographic setting~\cite{Giataganas:2021jbj}: we show in this article that even holographic entanglement entropy using a simple setup in five-dimensional gravity can be well approximated by large $D$ expansion. 
Second, apart from Einstein gravity, the large $D$ expansion may also be give a direct link to different kind of theories at low $D$~\cite{Soda:1993xc,Emparan:2013xia}. In particular, it turns out that high dimensional conformal theories are related to nearly critical low-dimensional theories, where criticality refers to the deconfinement transition~\cite{Gursoy:2015nza,Betzios:2017dol,Betzios:2018kwn}. The ultimate $D = \infty$ limit corresponds to exactly critical theory. This correspondence can be made explicit by using gauge/gravity duality, and carrying out dimensional reduction in the bulk gravity. That is, AdS$_D$ black holes are solutions to $D$-dimensional Einstein gravity with a cosmological constant. Dimensionally reducing~\cite{Gouteraux:2011qh,Gursoy:2015nza} these theories gives, say, five-dimensional Einstein dilaton gravity with an exponential potential for the dilaton, $V(\phi) \propto \exp(\alpha_D \phi)$~\cite{Kulkarni:2012re,Kulkarni:2012in}. Now one can make the link to criticality explicit: In general, for dilaton potentials having exponential asymptotics at large values of $\phi$, i.e., $V(\phi) \sim \exp(\alpha \phi)$ as $\phi \to \infty$, a critical value of $\alpha$ arises~\cite{Gursoy:2007er}. That is, for vacuum geometries with such potentials $\phi$ diverges in low energy (infrared, IR) region. If $\alpha>\alpha_c$, where the critical value depends on normalization conventions, the theory is IR-confined, and for $0<\alpha<\alpha_c$ it is IR-deconfined. The correspondence between the large-$D$ limit and criticality means that as $D \to \infty$, $\alpha_D$ approaches the critical value $\alpha_c$ from below.

This correspondence between the large-$D$ limit and near-criticality means that our results, which are derived for solutions to plain Einstein gravity at large $D$, also apply to general near-critical (say five-dimensional) geometries if one considers small black holes, so that the near-horizon region is described by the large-$\phi$ asymptotics of the dilaton potential. Often this means the low-temperature limit in the dual field theory. Interestingly, also holographic models of pure Yang-Mills theory are defined in terms of asymptotically exponential potentials that are nearly critical, including nearly critical but deconfined\footnote{Since pure Yang-Mills is confining, these models may sound like bad models for the theory. However, in these works the main focus is a simple reproduction of the thermodynamics of Yang-Mills and full QCD both above and below the phase transition or crossover temperature, which becomes complicated if confinement is imposed.} and nearly critical confined models~\cite{Gursoy:2007cb,Gursoy:2007er,Gubser:2008ny}. 

However, applications to holographic nearly critical theories are not limited to models of the Yang-Mills theory. A class of models~\cite{Gursoy:2010kw,Gursoy:2010jh} which display higher order deconfinement phase transitions, and may provide holographic duals to spin models, is even closer to the exactly critical case than the Yang-Mills models. The gravity duals of near-critical theories may also provide examples of violation of cosmic censorship~\cite{AragonesFontbote:2024lor} and even models for dark matter~\cite{Csaki:2021gfm,Fichet:2022ixi}.

Our work therefore provides an approach to entanglement which complements earlier 
interesting results pertaining to entanglement at criticality involving  conformal quantum critical points, quantum phase transitions, $c$-functions, black holes, QCD phase diagram and holography~\cite{Osborne:2002zz,Fradkin:2006mb,Cai:2012nm,Ling:2015dma,Knaute:2017lll,Fujita:2020qvp,Baggioli:2023ynu,Karan:2023hfk,Li:2025ksw}.

We also remark that the (nearly) critical geometry is in fact the so-called linear dilaton background, i.e., a well-studied special supergravity solution~\cite{Aharony:1998ub,Bertoldi:2009yi,Gursoy:2010jh,Fichet:2023xbu,Barbosa:2024pyn}, which appears in the context of little string theory and has been argued to be $\alpha'$ exact. The large $D$ geometries are also connected to two-dimensional string theory backgrounds~\cite{Emparan:2013xia,Gursoy:2021vpu}, signaling the emergence of a special two-dimensional conformal symmetry in the limit of large $D$. Note also that exponential potentials appear quite in general in string compactifications, so it is possible that nearly critical theories also appear as results of compactification of some other, more complex high dimensional backgrounds than the  high-$D$ Einstein gravity.

Large $D$ methods can also be applied to black hole gravity solutions having finite charge. In this context, an interesting region to study is the region of near-extremal black holes, which are dual to dense matter at low temperature. The horizon region of extremal black holes is given (in the time and holographic directions) by the AdS$_2$ geometry. The AdS$_2$ geometry is signals the presence of an $1+1$ dimensional CFT arising in the low energy limit, signaling the appearance of a quantum critical region~\cite{Liu:2009dm,Faulkner:2009wj,Iqbal:2011in}. This AdS$_2$ region is of particular interest in the context of holography for condensed matter, and is linked to the Sachdev-Ye-Kitaev models~\cite{Sarosi:2017ykf}. As it turns out, our method applies particularly nice to the AdS$_2$ case: unlike in other cases, the NH contributions to the entanglement entropy can be expressed in terms of elementary functions for the AdS$_2$ backgrounds.

The rest of the article is organized as follows. In Sec.~\ref{sec:setup} we introduce our setup, discuss dimensional reduction of the large $D$ picture to nearly critical five-dimensional gravity solutions, and show how the holographic entanglement entropy behaves under dimensional reduction. In Sec.~\ref{sec:neutralBH}, we apply the large $D$ method to study entanglement entropy in neutral black hole backgrounds. Secs.~\ref{sec:charged},~\ref{sec:extremal}, and~\ref{sec:soliton} present the generalizations of the analysis of Sec.~\ref{sec:neutralBH} to charged black holes, extremal black holes, and soliton geometries, respectively. In Sec.~\ref{sec:expansion} we study the expansion of the entanglement entropy at large size of the entanglement region and at large $D$, starting from strips and generalizing to other regions. Finally, in Sec.~\ref{sec:conclusions}, we conclude by summarizing the results and discussing future directions. The appendix contains details on the connection between exact (dimensionally reduced) AdS solutions and more general near-critical backgrounds.

\section{Entanglement entropy at criticality}  \label{sec:setup}

In this section, we will discuss our setup, starting from the $D$-dimensional gravity, with the understanding that $D$ will be taken to be large. We then discuss dimensional reduction to five dimensions, to make contact to 3+1 dimensional field theories.

We start from Einstein gravity in $D$ bulk dimensions,
\be \label{eq:highDS}
 S = \frac{1}{16\pi G_D} \int d^D z \sqrt{-\det G} \left[R-\Lambda\right] \ ,
\ee
which is interpreted as a gravity dual for a $d=D-1$ dimensional CFT. We denote the metric as
\be
 ds_D^2 = G_{MN}(z)dz^Mdz^N \ .
\ee

As discussed in the introduction, we will be working in high-dimensional geometries but we also want to make contact with field theories in 3+1 dimensions.
This link can be made concrete by dimensionally reducing the action~\eqref{eq:highDS} to a lower-dimensional Einstein-dilaton gravity, following~\cite{Gouteraux:2011qh}.  In order to discuss the setup for 3+1 dimensional field theories, we set the dimensionality of the lower-dimensional gravity to five.
We divide the coordinates $z^M$ into a subset of five first coordinates $x^0, x^1, \ldots x^4$ (which include the time $x^0\equiv t$ and the holographic coordinate $x^4 \equiv r$) and the remaining coordinates $y_1, y_2, \ldots y_{D-5}$. Then, assuming that the $D$-dimensional metric depends only on the coordinates $x^\mu$, we may parametrize
\be \label{eq:metricdecomp}
 ds_D^2 = G_{MN}(x)dz^Mdz^N = e^{\delta_1\phi(x)} ds_5^2 + e^{\delta_2\phi(x)}ds_{D-5}^2 
\ee
where we make the following choices:
\begin{align}
\begin{aligned}\label{eq:metricdefs}
        \delta_1 &= \frac{4}{3}\frac{\sqrt{d-4}}{\sqrt{d-1}} \ , &\qquad \delta_2 &= -\frac{4}{\sqrt{(d-4)(d-1)}} \ ,  &\\ 
       ds_5^2 &=  g_{\mu\nu}(x)dx^\mu dx^\nu \ , &\qquad ds_{D-5}^2 &=h_{ab}dy^ady^b \ . &
\end{aligned}
\end{align}
The choice of $\delta_1$ and $\delta_2$  is dictated by the consistency of the dimensional reduction as we discuss below.
We use here the notation that capital Latin indices $M, N, \ldots$ run over all dimensions, Greek indices $\mu, \nu, \ldots$ run over the five ``physical'' dimensions, and lowercase Latin indices $a, b, \ldots$  run over the remaining ``transverse'' directions.

The $D-5$ dimensional transverse manifold could have a curvature but we are interested in the case when it is flat. Actually, it is enough for us to take $h_{ab} = \delta_{ab}$. 
Integrating over the $y^a$ gives~\cite{Gouteraux:2011qh}
\be \label{eq:action5D}
 S = \frac{1}{16\pi G_5} \int d^5 x \sqrt{-\det g} \left[R-\frac{4}{3}g^{\mu\nu}\partial_\mu\phi\partial_\nu\phi -V(\phi)\right] 
\ee
where $G_5= G_D/V_\perp$ with $V_\perp$ the volume of the transverse manifold, and
\be \label{eq:Vreduced}
 V(\phi) = \Lambda \exp\left[\frac{4}{3}\sqrt{\frac{d-4}{d-1}}\,\phi\right] \ .
\ee
The ratio $\delta_1/\delta_2$ of the factors  in~\eqref{eq:metricdefs} was chosen such that linear terms in $\phi$ are absent in the reduced action, and their normalization  was chosen to produce the normalization coefficient $4/3$ of the kinetic term which turns out to be convenient in the Einstein frame action.

Note that for potentials of the form~\eqref{eq:Vreduced} the geometries are not asymptotically AdS$_5$ near the boundary where $\phi \to -\infty$. However, as argued in~\cite{Gursoy:2015nza,Betzios:2017dol,Betzios:2018kwn,Gursoy:2021vpu,Jarvinen:2024vdi}, the results are also relevant for a more general class of potentials with exponential asymptotics in the large coupling limit, e.g., 
\be\label{eq:Vlargephi}
 V(\phi) = V_0 e^{\alpha \phi}\left[1 + \morder{\frac{1}{\phi}}\right] \ , \qquad (\phi \to \infty) \ ,
\ee
with the parameter $\alpha$ close to the value $4/3$ obtained from~\eqref{eq:Vreduced} as $d \to \infty$.
In particular, the potential may be such that it admits asymptotically AdS$_5$ solutions, e.g. asymptotically as $\phi \to -\infty$ or arising from an extremum of the potential at finite $\phi$.
A typical configuration which essentially only depends on the asymptotic form of the potential is a small black hole, which usually means low temperature in field theory. We analyze the entanglement entropy (in the case of strips) for such geometries in Appendix~\ref{app:critical}, and argue that the results for the full geometry (i.e., geometry obtained with a generic potential having the asymptotics~\eqref{eq:Vlargephi}) indeed boil down to the analysis of the IR metric, which solves the Einstein-dilaton gravity with the exponential potential~\eqref{eq:Vreduced} and takes the form of the large-$D$ black hole dimensionally reduced to five dimensions. That is, when working with potentials of the generic form~\eqref{eq:Vlargephi}, ignoring subleading corrections at large $\phi$ is a  controlled approximation. In the following, we will restrict ourselves to purely exponential potentials, keeping in mind that they apply (with certain limitations, derived in the Appendix) also for the more general class of potentials that are only asymptotically exponential.

Let us then comment on the computation of the entanglement entropy. We consider a region $A$ in the $d$-dimensional CFT which has a nontrivial shape in the spatial directions $x^1$, $x^2$, and $x^3$, but extends over all the transverse directions $y^a$. According to~\cite{Ryu:2006bv,Ryu:2006ef}, entanglement entropy $S_E(A)$, obtained through reducing on the subsystem defined by the region $A$, is given by minimizing
\be \label{eq:AreahighD}
 \frac{1}{4G_D}\textrm{Area}[\Gamma] = \frac{1}{4G_D} \int_\Gamma d^{d-1}z\,\sqrt{-\det G_\Gamma}  
\ee
where $\Gamma$ is (any) bulk codimension two surface anchored to the boundary $\partial A$ of the region $A$, and $G_\Gamma$ is the induced metric on $\Gamma$. Because of our choice of $A$, the relevant surfaces $\Gamma$ extend trivially over the whole transverse space spanned by the coordinates $y^a$. For such surfaces, we observe that
\be
 \sqrt{-\det G_\Gamma} = e^{3\delta_1\phi+(d-4)\delta_2\phi}\sqrt{-\det g_{\widetilde{\Gamma}}}\sqrt{\det h} = \sqrt{-\det g_{\widetilde{\Gamma}}}\sqrt{\det h}
\ee
where $g_{\widetilde{\Gamma}}$ is the induced three-dimensional metric on the restriction of $\Gamma$ in lower dimensions, i.e., $\Gamma = \widetilde{\Gamma}\times \mathcal{M}_\perp$ where $\mathcal{M}_\perp$ is the transverse manifold spanned by $y^a$. Integrating over $y^a$ we find that 
\be \label{eq:Area5D}
 \frac{1}{4G_D}\textrm{Area}[\Gamma] = \frac{1}{4G_5}
 \int_{\widetilde{\Gamma}} d^3 x \sqrt{-\det g_{\widetilde{\Gamma}}}
\ee
where we inserted  $G_5= G_D/V_\perp$. That is, as expected, the entanglement entropy can be computed either in the $D$-dimensional setup or by using the dimensionally reduced (five-dimensional) setup.

Note that the area in~\eqref{eq:AreahighD} or~\eqref{eq:Area5D} 
is divergent near the boundary. We will regularize by adding a cutoff at a distance $\eps$ from the boundary, and by excluding all terms that are divergent as $\eps \to 0$.

We will be mostly focusing on the case where $A$ is a strip, e.g., a region with $0<x^1<L$, where $L$ is the length of the strip. As usual, the result for the entanglement entropy in this case may be written as a parametrization on the turning point in the holographic coordinate $r=x^4$ which we denote by $r_*$. Working in the five-dimensional picture, and assuming a homogeneous diagonal metric
\be \label{eq:5dgenmetric}
 ds_5^2 = g_{rr}(r)dr^2 - g_{tt}(r)dt^2 + g_{xx}(r)d\mathbf{x}^2 \ ,
\ee
where $\mathbf{x} = \{x^1,x^2,x^3\}$, we find the standard expressions
\begin{align} \label{eq:stripintegrals}
\begin{aligned}
 \textrm{Area}(r_*) &= 2 V_2\,\Bigg|\int_{r_\mathrm{bdry}}^{r_*}dr\,\frac{g_{xx}(r)g_{rr}(r)^{1/2}}{\sqrt{1-\frac{g_{xx}(r_*)^3}{g_{xx}(r)^3}}}\Bigg| \ , & \\ L(r_*) &= 2\, \Bigg|\int_{r_\mathrm{bdry}}^{r_*}dr\,\frac{g_{rr}(r)^{1/2}}{g_{xx}(r)^{1/2}\sqrt{\frac{g_{xx}(r)^3}{g_{xx}(r_*)^3}-1}}\Bigg| \ . &
\end{aligned}
\end{align}
Here we already extremized the area, $r_\mathrm{bdry}$ denotes the location of the boundary (including the cutoff which regularizes the divergence), and $V_2$ is the volume factor arising from integrating over $x^2$ and $x^3$. The absolute values in~\eqref{eq:stripintegrals} are needed because we did not fix the gauge yet, so it is not clear which of the  bounds $r_*$ and $r_\mathrm{bdry}$ is larger.

Before proceeding, let us point out that our setup has simple scaling symmetries. First, there is a symmetry linked to the AdS scale $\ell$, given by the mappings
\be
 \Lambda \mapsto \frac{\Lambda}{\mu_\ell^2} \ , \qquad G_{MN} \mapsto \mu_\ell^2 G_{MN} \ , \qquad G_D \mapsto \mu_\ell^{D-2} G_D 
\ee
in the $D$-dimensional setup. Indeed, the action of~\eqref{eq:highDS} is invariant under this transformation. The dimensionally reduced version of the mapping is
\be
 \Lambda \mapsto \frac{\Lambda}{\mu_\ell^2} \ , \!\!\qquad g_{\mu\nu} \mapsto \mu_\ell^{\frac{2}{3}(d-1)} g_{\mu\nu} \ , \!\!\qquad \phi \mapsto \phi - \frac{\sqrt{(d-4)(d-1)}}{2}\log \mu_\ell \ , \!\!\qquad G_5 \mapsto \mu_\ell^{d-1} G_5
\ee
which leaves~\eqref{eq:action5D} invariant.

The second scaling symmetry affects all coordinates in the system, therefore changing also the scale of the boundary theory. In the high-dimensional picture, we may write simply
\be
  z_M \mapsto \mu_s z_M \ , \qquad  \qquad G_{MN} \mapsto \frac{1}{\mu_s^2} G_{MN}
\ee
so that the line element $G_{MN}dz^M dz^N$ is invariant. Consequently, the Ricci scalar $R$ and the integration measure $d^Dz\sqrt{-\det G}$ are invariant as well. The five-dimensional counterpart of this transformation is
\be
 x_\mu \mapsto \mu_s x_\mu \ , \qquad g_{\mu\nu} \mapsto \frac{1}{\mu_s^2} g_{\mu\nu} \ .
\ee

\section{Neutral black hole backgrounds} \label{sec:neutralBH}

We start our analysis from the (planar) black hole solutions. The high-dimensional solution to the gravity defined by the action~\eqref{eq:highDS} reads
\be \label{eq:metricD}
 ds_D^2 = \frac{\ell^2}{r^2}\left(\frac{dr^2}{f(r)} - f(r)dt^2 + d\mathbf{z}^2\right)
\ee
where we set $\Lambda = -d(d-1)/\ell^2$, the spatial coordinates are $z_1, z_2, \ldots z_{d-1}$, and
\be \label{eq:fexp}
 f(r) = 1 -\left(\frac{r}{r_h}\right)^d
\ee
with $r=r_h$ being the location of the black hole horizon. The dimensionally reduced metric reads in this case
\be \label{eq:metric5}
 ds_5^2 = \left(\frac{\ell}{r}\right)^{\frac{2}{3}(d-1)}\left(\frac{dr^2}{f(r)}-f(r) dt^2 + d\mathbf{x}^2\right) \ .
\ee
The solution for the dilaton is
\be \label{eq:phisol}
 \phi = \frac{1}{2}\sqrt{(d-4)(d-1)} \, \log \frac{r}{\ell}
\ee
but we will not need it in the analysis.

At large $d$, the blackening factor~\eqref{eq:fexp} reflects the expected membrane picture: $f \approx 1$ almost everywhere in space, so that deviation from the vacuum geometry is only found at small distances $r_h-r \sim 1/d$ away from the horizon. This picture suggests an approach, where one studies separately the geometry near the horizon, uses vacuum solutions elsewhere, and by combining the result obtains an accurate solution for the whole geometry. This approach has been successful in the past among other things in the study of quasi normal modes (see, e.g.,~\cite{Emparan:2020inr}). Here we will apply it to the entanglement entropy.

The basic idea of the approach can be demonstrate by explicitly considering the metric at large $d$. 
The near-boundary (NB) limit of the metric is obtained by taking $d \to \infty$ at fixed value of coordinate $r$. This gives the vacuum AdS geometry:
\begin{align}
\begin{aligned}\label{eq:NBmetric}
 ds_D^2 &= \frac{\ell^2}{r^2}\left(dr^2 - dt^2 + d\mathbf{z}^2\right)\left[1+\morder{\left(\frac{r}{r_h}\right)^d}\right] & \\
  ds_5^2 &= \left(\frac{\ell}{r}\right)^{\frac{2}{3}(d-1)}\left(dr^2-dt^2 + d\mathbf{x}^2\right)\left[1+\morder{\left(\frac{r}{r_h}\right)^d}\right] &
\end{aligned}
\end{align}
where the corrections due to the nontrivial blackening factor may be treated perturbatively. 

To study the near-horizon (NH) limit, we first define a new variable
\be \label{eq:wdef}
 w \equiv \left(\frac{r}{r_h}\right)^d
\ee
and take the limit $d \to \infty$ keeping $w$ fixed instead of $r$. This means that 
$r = r_h w^{1/d} = r_h(1+(\log w)/d + \cdots)$ will be close to the horizon $r_h$ with corrections $\sim 1/d$. In this limit, we find that
\begin{align}
\begin{aligned}\label{eq:NHmetric}
  ds_D^2 &= \frac{\ell^2}{r_h^2}\left(\frac{r_h^2}{d^2w^2}\frac{dw^2}{f(w)} - f(w)dt^2 + d\mathbf{z}^2\right)\left[1+\morder{\frac{1}{d}\log w}\right] & \\
  ds_5^2 &= \left(\frac{\ell}{r_h}\right)^{\frac{2}{3}(d-1)} w^{-2/3}\left(\frac{r_h^2}{d^2w^2}\frac{dw^2}{f(w)}-f(w)dt^2 + d\mathbf{x}^2\right)\left[1+\morder{\frac{1}{d}\log w}\right] &
\end{aligned}
\end{align}
where the blackening factor simplifies to
\be \label{eq:fNH}
 f(w) = 1-w \ .
\ee

Now we note that the range of validity of the NB form of the metric in~\eqref{eq:NBmetric} is
\be
 r_h-r \gg \frac{1}{d} \ , \qquad w \ll 1 
\ee
in the $r$ and $w$ coordinates, respectively, whereas the NH form~\eqref{eq:NHmetric} is valid when
\be
 r_h-r \ll 1 \ , \qquad w \gg e^{-d} \ .
\ee
Therefore, there is a range of coordinates,
\be \label{eq:matchingrange}
\frac{1}{d} \ll r_h-r \ll 1 \ , \qquad e^{-d} \ll w \ll 1 \ ,
\ee
where both the NB and NH expressions are good approximations, and this range grows with increasing $d$. This demonstrates that the approach, where one analyzes observables separately in the near boundary and near horizon region and matches in the middle, will become precise in the limit of large $d$. This is a well-known method which applies to different observables, (in particular the computation of the fluctuation spectra), in large-$D$ gravity~\cite{Emparan:2020inr}. We now apply this idea to embeddings needed in the computation of entanglement entropy.

\subsection{Entanglement entropy near the boundary: direct expansion}

We start by analyzing the embeddings near the boundary. While we will a more complicated embeddings for the final construction that combines expressions from both the near boundary and near horizon regions, we will start by embeddings which are close enough to the boundary that the near-horizon part is not needed.

The expressions for the case of the empty AdS metric~\eqref{eq:NBmetric} are well known, see, e.g.~\cite{Ryu:2006ef}. However, it is also possible to write an improved approximation to the near-boundary embeddings by treating the inclusion of the blackening factor as a perturbation (see~\cite{Giataganas:2021jbj}). First, inserting the exact geometry~\eqref{eq:metric5} in~\eqref{eq:stripintegrals} gives
\begin{align} \label{eq:BHexact}
\begin{aligned}
 \textrm{Area}(r_*) &= 2 V_2\ell^{d-1}\,\int_{\epsilon}^{r_*}dr\,\frac{1}{r^{d-1}\sqrt{f(r)}
 \sqrt{1-\frac{r^{2d-2}}{r_*^{2d-2}}}} \ , & \\ L(r_*) &= 2\, \int_{\epsilon}^{r_*}dr\,\frac{1}{\sqrt{f(r)}
 \sqrt{\frac{r_*^{2d-2}}{r^{2d-2}}-1}} \ , &
\end{aligned}
\end{align}
where $f(r)$ is given in~\eqref{eq:BHexact} and we set $r_\mathrm{bdry}=\epsilon \ll 1$ as the boundary is located at $r=0$ in the chosen coordinates. As usual, the area integral here is divergent. Subtracting the divergence, i.e., the term $2V_2\ell^{d-1}/((d-2)\epsilon^{d-2})$ in the area, we find
\begin{align} \label{eq:BHexactreg}
\begin{aligned}
 \textrm{Area}_\textrm{reg}(r_*) &= 2 V_2\ell^{d-1}\,
 \left\{\int_{0}^{r_*}dr\,\frac{1}{r^{d-1}}
 \left[\frac{1}{\sqrt{f(r)}
 \sqrt{1-\frac{r^{2d-2}}{r_*^{2d-2}}}}-1\right]-\frac{1}{(d-2)r_*^{d-2}}
 \right\} \ , 
 & \\ L(r_*) &= 2\, \int_{0}^{r_*}dr\,\frac{1}{\sqrt{f(r)}
 \sqrt{\frac{r_*^{2d-2}}{r^{2d-2}}-1}} \ , &
\end{aligned}
\end{align}
where we could take $\epsilon \to 0$ as the integrals are now convergent. We will omit the subscript ``reg'' below as all area integrals will be regulated in the same way.

Now the desired analytic expressions are obtained by developing the blackening factor in the integrands as a series in $r/r_h$.  The procedure leads to 
\begin{align} \label{eq:BHexactser}
\begin{aligned}
 \textrm{Area}(r_*) &= \frac{2\pi V_2\ell^{d-1}}{r_*^{d-2}} \sum _{k=0}^{\infty} \frac{  (-1)^k  \Gamma \left(\frac{k+1}{2}+\frac{k+1}{2 (d-1)}\right)}{\left[d (k-1)+2\right] \Gamma \left(\frac{1}{2}-k\right) \Gamma (k+1) \Gamma \left(\frac{k}{2}+\frac{k+1}{2 (d-1)}\right)}\left(\frac{r_*}{r_h}\right)^{kd}
 \,
 \ ,  & \\ 
 L(r_*) &= \frac{2\pi r_*}{d}\, \sum _{k=0}^{\infty} \frac{  (-1)^k  \Gamma \left(\frac{k+3}{2}+\frac{k+1}{2 (d-1)}\right)}{\Gamma \left(\frac{1}{2}-k\right) \Gamma (k+2) \Gamma \left(\frac{k+2}{2}+\frac{k+1}{2 (d-1)}\right)}\left(\frac{r_*}{r_h}\right)^{kd} \ . &
\end{aligned}
\end{align}
Note that the regulator terms of the area integral in~\eqref{eq:BHexactreg} only contribute to the terms with $k=0$ in the series.

\begin{figure}[hbt!]
\centering
\includegraphics[width=0.47\textwidth]{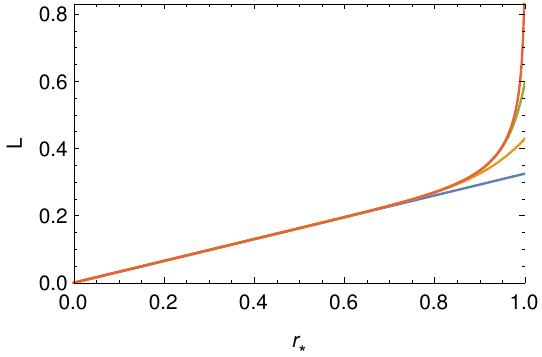}\hspace{4mm}%
\includegraphics[width=0.49\textwidth]{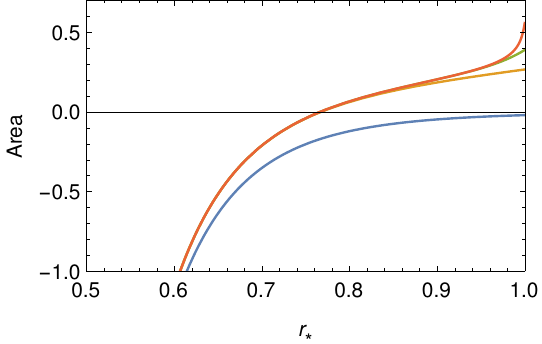}
\vspace{3mm}

\includegraphics[width=0.7\textwidth]{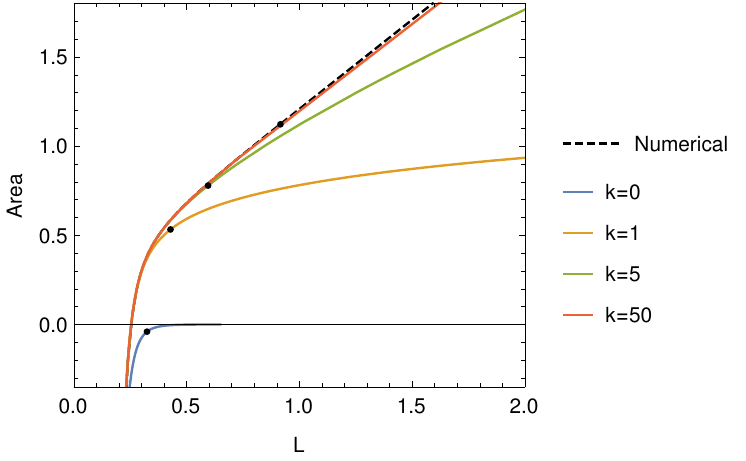}
\caption{Entanglement entropy for strips with $d=10$ (i.e., AdS$_{11}$ black holes) using the direct expansion~\protect\eqref{eq:BHexactser}. Top left: Length of the strip as a function of the turning point $r_*$. Top right: Regulated area of the embedding as a function of $r_*$. Bottom: Area as a function of the length.}
\label{fig:near_brdy_exp}
\end{figure}

We compare the results from these series at $d=10$, including terms up to $k=0$, $1$, $5$, and $50$ in Fig.~\ref{fig:near_brdy_exp}, {shown as the solid curves}. In these plots we set $r_h=V_2=\ell=1$. Both the area and the length diverge for $r_* \to r_h$ which sets the convergence radius of the expansion (see panels in the top row). { The bottom panel shows the comparison between our analytical and numerical results. The exact result, given as the dashed black curve, is obtained by numerically integrating (following~\cite{Jain:2022csf}) Eqs.~\eqref{eq:BHexactreg}. Even if convergence with increasing $d$ is expected to improve, including only a few terms does not give a good approximation of the result $\textrm{Area}(L)$ at all $L$. 
However, including terms up to $k=50$, we see that the expected transition from conformal behavior ($\textrm{Area}\propto L^{-d+2}$ for narrow strips) to linear behavior ($\textrm{Area} \propto L$ for wide strips) is well reproduced.

Note that the exact result for both the area and the length diverges at $r_* = r_h$, while the truncated results stay finite. That is, the truncation error also diverges at this point. Therefore, beyond this point (i.e., when $r_*>r_h$), the truncated results are definitely unreliable. We show the point  $r_* = r_h$ for the truncated results as the black dots in the bottom panel.
As one can see, these black dots indeed mark approximately the onset of the failure of the analytical results to match the numerical version: starting from the vicinity of the dots, the approximation starts being unreliable as $L$ (and therefore $r_*$) grows.
}

Finally we also remark that the factor in the square brackets in the expression for the area in~\eqref{eq:BHexactser}, i.e., $d(k-1)+2$ is of higher order in $1/d$ for $k=1$ than for any other value of $k$. This enhances the $k=1$ term with respect to all the other terms at large $d$. This is readily visible in Fig.~\ref{fig:near_brdy_exp} as the expression with $k=1$ included (orange curve) is a significantly better approximation than the result with the $k=0$ term only, whereas adding the other higher order terms in $k$ has much milder effect. We give another interpretation to this observation in Sec.~\ref{sec:expansion}.

\begin{figure}[hbt!]
\centering
\includegraphics[width=0.6\textwidth]{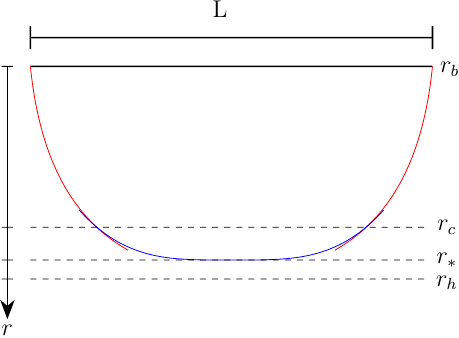}
\caption{Sketch of the large $D$ method.}
\label{fig:method_sketch}
\end{figure}

\subsection{Minimal surfaces near the boundary and near the horizon}

We then proceed to the approach where we use the geometry~\eqref{eq:NBmetric} to compute the minimal surface near the boundary, the geometry~\eqref{eq:NHmetric} to compute the surface near the horizon, and match in the middle. First we need the NB expressions for the area and the length, but integrated to a matching point $r=r_c$ within the interval~\eqref{eq:matchingrange} instead of $r=r_*$ (see the sketch in Fig.~\ref{fig:method_sketch}). Moreover, instead of directly using the simplest approximation of the NB geometry, i.e.~\eqref{eq:NBmetric}, we can treat the effect of the blackening factor perturbatively as above. That is, we start from
\begin{align} \label{eq:BHgen}
\begin{aligned}
 \textrm{Area}(r_*,r_c) &= 2 V_2\ell^{d-1}\,
 \left\{\int_{0}^{r_c}dr\,\frac{1}{r^{d-1}}
 \left[\frac{1}{\sqrt{1-\left(\frac{r}{r_h}\right)^d}\sqrt{1-\frac{r^{2d-2}}{r_*^{2d-2}}}}-1\right]-\frac{1}{(d-2)r_c^{d-2}}
 \right\} \ , 
 & \\ L(r_*,r_c) &= 2\, \int_{0}^{r_c}dr\,\frac{1}{\sqrt{1-\left(\frac{r}{r_h}\right)^d}\sqrt{\frac{r_*^{2d-2}}{r^{2d-2}}-1}} \ . &
\end{aligned}
\end{align}
Now expanding the blackening factors and integrating gives the NB approximation
\begin{align} \label{eq:NBneutral}
\begin{aligned}
 A_\nb(r_*,r_c) &= \frac{2\sqrt{\pi } V_2\ell^{d-1}}{r_c^{d-2}}\,\sum_{k=0}^{k_c}
 \frac{ (-1)^k \, _2F_1\left(\frac{1}{2},\frac{d (k-1)+2}{2 (d-1)};\frac{d (k+1)}{2 (d-1)};\left(\frac{r_c}{r_*}\right)^{2 d-2}\right)}{\left[d (k-1)+2\right] \Gamma \left(\frac{1}{2}-k\right) \Gamma (k+1)}\left(\frac{r_c}{r_h}\right)^{kd} \ , 
 & \\ L_\nb(r_*,r_c) &= \frac{2\sqrt{\pi }r_c^d}{dr_*^{d-1}}\, \sum_{k=0}^{k_c}\frac{ (-1)^k \, _2F_1\left(\frac{1}{2},\frac{d (k+1)}{2 (d-1)};\frac{d (k+3)-2}{2 (d-1)};\left(\frac{r_c}{r_*}\right)^{2 d-2}\right)}{ \Gamma \left(\frac{1}{2}-k\right) \Gamma (k+2)}\left(\frac{r_c}{r_h}\right)^{kd} \ , &
\end{aligned}
\end{align}
where we included corrections only up to $k=k_c$.

In order to write down the complete expressions we need similar approximations for the NH geometry of~\eqref{eq:NHmetric}. Using~\eqref{eq:stripintegrals} we find that the integrals we need to compute in this case are
\begin{align} \label{eq:NHintegrals}
\begin{aligned}
 A_\nh(w_*,w_c) &= 2 V_2\left(\frac{\ell}{r_h}\right)^{d-1}\frac{r_h}{d}\,
 \int_{w_c}^{w_*}dw\frac{1}{ w^2\sqrt{f(w)} \sqrt{1-\frac{w^2}{w_*^2}}}\, \ , 
 & \\ 
 L_\nh(w_*,w_c) &= \frac{2r_h}{d}\, \int_{w_c}^{w_*}dw\,\frac{1}{ w\sqrt{f(w)} \sqrt{\frac{w_*^2}{w^2}-1}} \ , &
\end{aligned}
\end{align}
where $w_c= r_c^d$, $f(w)=1-w$, and $w_*$ is the turning point of the surface.
These integrals can be expressed in terms of elliptic functions as follows:
\begin{align} \label{eq:NHneutral}
\begin{aligned}
 A_\nh(w_*,w_c) &= \frac{2V_2 r_h}{d w_* \sqrt{1+w_*}} \left(\frac{\ell}{r_h}\right)^{d-1} \Bigg[\frac{1}{w_c}\sqrt{\frac{(1+w_*) (w_*-w_c) (w_*+w_c)}{1-w_c}}+&\\
 &\ +\left(1+w_*^2\right) F\left(\varphi |m\right)-(1+w_*) E\left(\varphi |m\right)+w_*(1-w_*)  \Pi \left(\frac{2}{1+w_*};\varphi \Big|m\right)\!\Bigg] \, , 
 & \\ 
 L_\nh(w_*,w_c) &= \frac{4 r_h}{d\sqrt{1+w_*}} F\left(\varphi |m\right)=\frac{4r_h\, \text{dn}^{-1}\left(\sqrt{\frac{1-w_*}{1-w_c}}\big|m\right)}{d \sqrt{1+w_*}} \ , &
\end{aligned}
\end{align}
where the angle $\varphi$ and the parameter $m$ are given by
\be
 \varphi = \sin ^{-1}\left(\sqrt{\frac{\left(1+w_*\right) (w_*-w_c)}{2w_*(1-w_c)}}\right) \ , \qquad m = \frac{2 w_*}{1+w_*} \ ,
\ee
respectively. Here $F$, $E$, and $\Pi$ are the (incomplete) elliptic integrals of the first, second, and third kind, respectively, and $\text{dn}^{-1}$ is the inverse of one of the Jacobi elliptic functions, the delta amplitude.

\subsection{Entanglement entropy via matching}

The full expressions for the length and area are obtained by combining the NB~\eqref{eq:NBneutral} and NH~\eqref{eq:NHneutral} results. However to do this, we need to specify two things. First, while we know that $r_c$ needs to lie within the range~\eqref{eq:matchingrange}, we should try to pick an optimal value within this range. Second, the turning point $r_*$ appears as an integration constant in our NB expressions but since it is not reached in the NB region  (assuming $r_c<r_*$ which is the domain where our method nontrivially combines information both from the NB and NH regions), it is not clear what its value and interpretation is. We expect that it is related to the turning point in the NH region, denoted by $w_*$ above, but the precise form of the relation is so far unclear. As we show now, both the parameters $r_c$ and $r_*$ can be determined by studying matching between the NB and NH geometries.

Let us first discuss the relation between $r_*$ and $w_*$. This is most conveniently read by comparing the derivatives $dx_1/dr$ for the embedding, which can be read off from the expressions for the length in~\eqref{eq:BHgen} and in~\eqref{eq:NHintegrals}. We find that
\be \label{eq:derexps}
 \left.\frac{dx_1}{dr}\right|_\mathrm{NB} = \frac{1}{\sqrt{\frac{r_*^{2d-2}}{r^{2d-2}}-1}} \ , \qquad \left.\frac{dx_1}{dr}\right|_\mathrm{NH} = d \frac{r^{d-1}}{r_h^d}\frac{r_h}{d}\frac{1}{w \sqrt{1-w}\sqrt{\frac{w_*^{2}}{w^{2}}-1}} \ ,
\ee
where we restricted to the leading NB expression and included the Jacobian in the NH expression. In the matching region~\eqref{eq:matchingrange}, both NB and NH approximations should work, and both expressions should be good approximations in the limit of large $d$. Therefore, in particular, if apply both the NB and NH approximations at the same time, the two expressions should match. To implement this, we write both expressions in~\eqref{eq:derexps} in terms of $w$, approximate $w^{1/d}\approx 1$, and drop corrections suppressed by powers of $w$. This gives
\be
 \left.\frac{dx_1}{dr}\right|_\mathrm{NB} \approx \frac{1}{\sqrt{\frac{r_*^{2d-2}}{r_h^{2d-2}w^2}-1}}\ , \qquad \left.\frac{dx_1}{dr}\right|_\mathrm{NH} \approx \frac{1}{\sqrt{\frac{w_*^{2}}{w^{2}}-1}} \ .
\ee
Therefore matching the expressions gives $w_*= r_*^{d-1}/r_h^{d-1}$. That is, taking only the leading behavior at large $d$, we find 
\be \label{eq:srel}
 w_* = \left(\frac{r_*}{r_h}\right)^d\ ,
\ee
which simply follows the definition~\eqref{eq:wdef}. However, this is not always true: in Sec.~\ref{sec:soliton} we show that for another geometry, the matching result differs from the coordinate definition.

The optimal value for $r_c$ is found by using a simpler argument. The error from using the NB expansion~\eqref{eq:NBneutral} with a cutoff is $\sim (r_c/r_h)^{(k_c+1)d}$, while the error in the NH approximation is $\sim (\log w_c)/d \sim \log (r_c/r_h)$. The error is minimized when these expressions are of the same order. Requiring this, and solving iteratively, gives
\be
 \frac{r_c}{r_h} = 1 - \frac{1}{(k_c+1)d}\log d +\morder{\frac{1}{d}} \ .
\ee
Therefore, we will use 
\be
 \frac{r_c}{r_h} = 1 - \frac{1}{(k_c+1)d}(\log d +C) \ .
\ee
As for the parameter $C$, one can test several values, and pick the one that produces smoothest behavior of the integrals when $r_*$ crosses $r_c$. In addition, we will need to choose a value for the cutoff $k_c$ in~\eqref{eq:NBneutral}. Below we will use $k_c=1$, motivated by the observation that the $k=1$ terms in~\eqref{eq:BHexactser} and~\eqref{eq:NBneutral} are enhanced at large $d$, and $C=-1$.

%    \item Discuss how the gluing point $r_c$ can be determined and how the optimal choice scales with D

Collecting the results, the final matched expressions for the length and the area are
\begin{align}\label{eq:neutralfinal}
\begin{aligned}
 \mathrm{Area}(r_*) &= \left\{
 \begin{array}{lc}
    A_\nb(r_*,r_*)\ ,  & \qquad (r_* \le r_c)  \\
    A_\nb(r_*,r_c) +A_\nh(r_*^d,r_c^d)\ , &  \qquad (r_* > r_c) 
 \end{array}
 \right. & \\
 L(r_*) &= \left\{
 \begin{array}{lc}
    L_\nb(r_*,r_*)\ ,  & \qquad (r_* \le r_c)  \\
    L_\nb(r_*,r_c) +L_\nh(r_*^d,r_c^d)\ , &  \qquad (r_* > r_c) 
 \end{array}
 \right. & 
\end{aligned}
\end{align}
with the functions given in~\eqref{eq:NBneutral} and~\eqref{eq:NHneutral}.
For definiteness, we may set $k_c=1$ and $r_c/r_h = 1 - \frac{1}{2d}(\log d-1)$.

We stress that the expressions~\eqref{eq:neutralfinal} were derived by assuming the large $d$ limit. However, while the $d$-dependence in the NH expressions in~\eqref{eq:NHneutral} has scaled out, the NB expressions~\eqref{eq:NBneutral} have complicated dependence on $d$. Therefore it is tempting to simplify the NB expressions further by dropping corrections suppressed at large $d$, for example in the arguments of the hypergeometric functions. However it turns out that this works for the expressions for length but not for the expressions for area: attempts to simplify the area formulas lead to expressions which work significantly worse in particular at low values of $L$. This is because the near boundary behavior and the renormalization process is sensitive to the value of $d$, with cancellations between the various terms. For example, naively dropping all subleading $1/d$ corrections in the $k=0$ term of~\eqref{eq:NBneutral} leads to heavily modified behavior: the area no longer approaches $-\infty$ in the limit of short strips.

\begin{figure}[hbt!]
\centering
\includegraphics[width=0.7\textwidth]{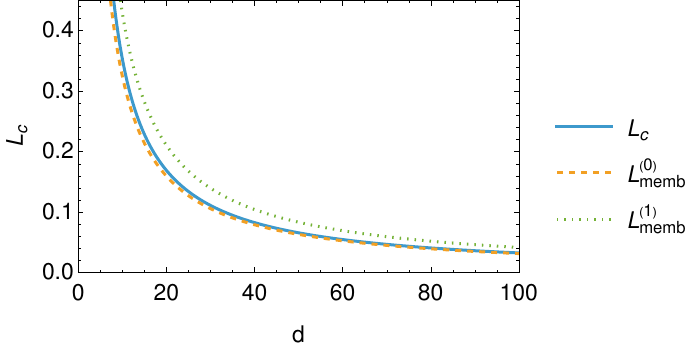}
\caption{The critical length $L_c$ at which near-horizon geometry starts to contribute as a function of $d$, compared to an approximation obtained by treating the black hole as a membrane, Eqs.~\eqref{eq:Lmemb0} and~\eqref{eq:Lmemb1}.}
\label{fig:critical_L}
\end{figure}

\subsection{Results}

Let us then analyze numerically the implications of the formulas~\eqref{eq:neutralfinal}. Thanks to the scaling symmetries discussed at the end of Sec.~\ref{sec:setup}, we may set $\ell=1$ and $r_h=1$ without loss of generality. We will also set the trivial transverse volume factor $V_2$ to one.

We start by checking the critical length of the strip at which the NH geometry starts to contribute significantly to the result. This is estimated by setting that $r_*=r_c$, i.e., we define the critical length as $L_c \equiv L(r_c)$. The result is shown as a function of $d$ as the blue curve in Fig.~\ref{fig:critical_L}. The length should be interpreted to be given in units of $r_h$. We compare the estimate to a simple ``membrane approximation'', obtained by replacing the black hole by a membrane at $r=r_h$, and computing the length at which the embedding (in empty AdS) reaches the membrane. This length can be found analytically and in terms of the expressions listed above, see~\eqref{eq:BHexactser} and~\eqref{eq:NBneutral}, it is given by
\begin{equation} \label{eq:Lmemb0}
    L_\mathrm{memb}^{(0)} \equiv \left.L_\nb(r_h,r_h)\right|_{k_c=0} = \frac{2 \sqrt{\pi }\,  \Gamma \left(\frac{3 d-2}{2 (d-1)}\right)}{d\, \Gamma \left(\frac{2 d-1}{2 (d-1)}\right)} r_h \ .
\end{equation}
For comparison, we also include the same formula but with the leading correction due to the blackening factor,
\begin{equation} \label{eq:Lmemb1}
    L_\mathrm{memb}^{(1)} \equiv \left.L_\nb(r_h,r_h)\right|_{k_c=1} = \frac{2 \sqrt{\pi }\,  \Gamma \left(\frac{3 d-2}{2 (d-1)}\right)}{d\, \Gamma \left(\frac{2 d-1}{2 (d-1)}\right)} r_h +\frac{\sqrt{\pi }  \, \Gamma \left(\frac{2 d-1}{d-1}\right)}{2 d \, \Gamma \left(\frac{3 d-1}{2 (d-1)}\right)} r_h\ .
\end{equation}
These expressions are compared to the critical length $L_c$ in Fig.~\ref{fig:critical_L}. Interestingly, the simplest, empty AdS formula gives a good approximation to $L_c$. Note that all the lengths decay as $\sim 1/d$ at large $d$.

\begin{figure}[hbt!]
\centering
\includegraphics[width=0.48\textwidth]{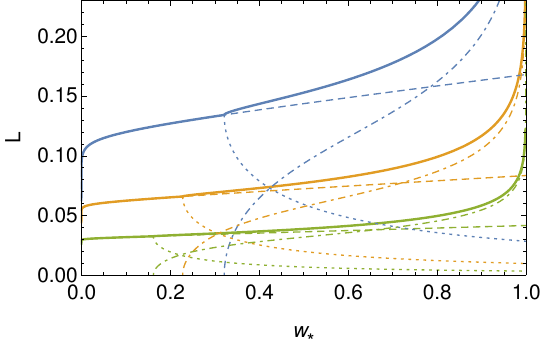}\hspace{4mm}%
\includegraphics[width=0.46\textwidth]{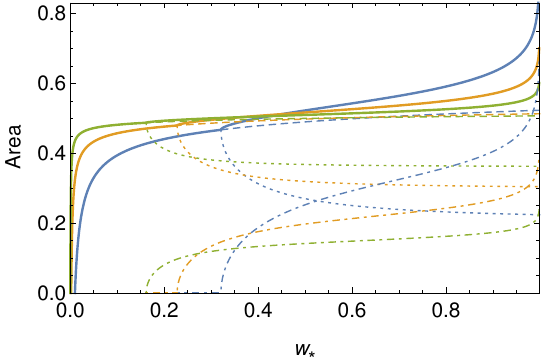}
\vspace{3mm}

\includegraphics[width=0.8\textwidth]{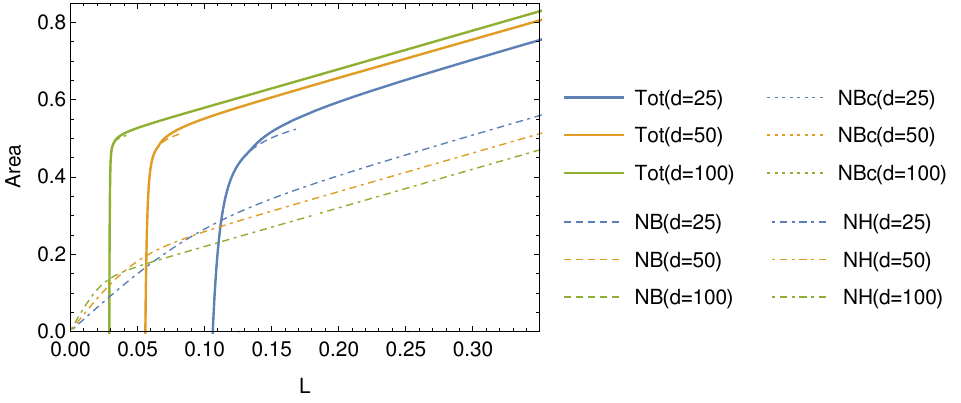}
\caption{Breakdown of the analytical approximation of the Entanglement entropy to near-boundary and near-horizon contributions for neutral black holes. Top left: Length of the strip as a function of the turning point $w_*$. Top right: Regulated area of the embedding as a function of $w_*$. Bottom: Area as a function of the length.  We use $d=25$, $50$, and $100$ shown as the blue, orange, and green curves, respectively, as indicated in the Legend. The thick solid curves show the full result, labeled as  ``Tot''. The dotted (dotdashed) curves show the NB (NH) contribution to the full result, labeled as ``NBc'' (``NH''). Finally, the dashed curve shows the result if the NH terms are ignored and only the NB expression is used (labeled as ``NB'').    }
\label{fig:breakdown_neutral}
\end{figure}

We show the entanglement entropy given by the analytic result~\eqref{eq:neutralfinal} in Fig.~\ref{fig:breakdown_neutral}. Apart from simply plotting the result as a function of the strip length (bottom row), we show the length and the area separately as a function of the turning point on the top row. We chose to use $w_* = (r_*/r_h)^d$ instead of $r_*$ because this makes details near the horizon better visible. In addition to the final result (thick curves), we show its breakdown to NB and NH component (dotted and dotdashed curves, respectively) as well as the result obtained by using the truncated NB series only (dashed curves). Unsurprisingly, the matching between the NB and NH geometries, represented by the kinks in the curves, becomes smoother as $d$ increases and the analytic approximation becomes more accurate. Note also that the kinks as a function of $w_*$ are somewhat more pronounced than in the final plot of the area as a function of the length.

\begin{figure}[hbt!]
\centering
\includegraphics[width=0.7\linewidth]{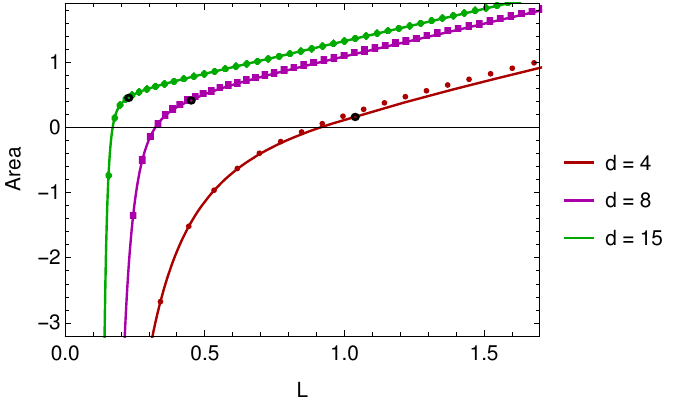}
\caption{Comparison of the analytic expressions (curves) to exact numerical results (data points) at relatively low $d$. The black circles denote the transition of the turning point of the embedding from the near-boundary to the near-horizon region. 
}
\label{fig:large_D_comp}
\end{figure}

Finally we compare the analytic formula to the exact result (that can be obtained numerically at low $d$) in Fig.~\ref{fig:large_D_comp}. In this plot, the curves show the analytic large-$d$ approximation and dots are numerical data for the exact result. The black circles show the points where we switch from using the NB formula only to using a combination of NB and NH results. The kinks at these points are less clearly visible than in Fig.~\ref{fig:breakdown_neutral} because we have zoomed out. Remarkably, the analytic result is a good approximation even for $d=4$, i.e., the $\mathcal{N}=4$ Super-Yang-Mills. As $d$ increases, the approximation becomes better.

\section{Charged backgrounds}\label{sec:charged} 

As it turns out, our analysis has a straightforward generalization to charged backgrounds. We first generalize~\eqref{eq:highDS} by adding a Maxwell term: 
\be \label{eq:highDSMaxwell}
 S = \frac{1}{16\pi G_D} \int d^D z \sqrt{-\det G} \left[R-\Lambda- \frac{1}{4} F_{MN} F^{MN}\right] \ ,
\ee
where $F_{MN} = \partial_M A_N-\partial_N A_M$ and $A_M$ is a gauge field.
Now reducing this to five dimensions as in Sec.~\ref{sec:setup} (setting the extra components of the gauge field to zero and assuming that the rest depend only on the first five coordinates), the five dimensional action~\eqref{eq:action5D} receives an extra term~\cite{Gouteraux:2011qh,Gursoy:2021vpu}:
\be \label{eq:action5DMaxwell}
 S = \frac{1}{16\pi G_5} \int d^5 x \sqrt{-\det g} \left[R-\frac{4}{3}g^{\mu\nu}\partial_\mu\phi\partial_\nu\phi -V(\phi) -\frac{1}{4}Z(\phi)F_{\mu\nu}F^{\mu\nu}\right] 
\ee
with 
\be \label{eq:Zreduced}
 Z(\phi) = \exp\left[-\frac{4}{3}\sqrt{\frac{d-4}{d-1}}\,\phi\right] \ .
\ee
That is, the coupling function $Z(\phi)$ of the gauge field is the inverse of the potential $V(\phi)$ given in~\eqref{eq:Vreduced}.

We are mostly interested in the charged black hole solution. In $D=d+1$ dimensions, the Reissner--Nordstr\"om black hole geometry is
\be \label{eq:RNmetricD}
 ds_D^2 = \frac{\ell^2}{r^2}\left(\frac{dr^2}{f_q(r)} - f_q(r)dt^2 + d\mathbf{z}^2\right)
\ee
with
\be \label{eq:fexpq}
 f_q(r) = 1 -\left(1+\frac{dq}{d-2}\right)\left(\frac{r}{r_h}\right)^d + \frac{dq}{d-2}\left(\frac{r}{r_h}\right)^{2d-2} \ ,
\ee
where the (squared) charge $q$ was normalized such that $0 \le q \le 1$ so that $q \to 1$ is the limit of an extremal black hole.
This geometry is supported by a gauge field
\be
 A_t = \mu\left[1-\left(\frac{r}{r_h}\right)^{d-2}\right]
\ee
with 
\be 
 \mu = \frac{\sqrt{2d(d-1)}\ell}{(d-2)r_h}\sqrt{q}\ .
\ee
Reducing to five dimensions, the geometry takes the same form as in Sec.~\ref{sec:neutralBH} except the blackening factor is modified,
\be \label{eq:metric5RN}
 ds_5^2 = \left(\frac{\ell}{r}\right)^{\frac{2}{3}(d-1)}\left(\frac{dr^2}{f_q(r)}-f_q(r) dt^2 + d\mathbf{x}^2\right) \ .
\ee
The dilaton profile is likewise the same as before, given in Eq.~\eqref{eq:phisol}. For later use we also recall the expressions for the temperature and the entropy density for this black hole. They are given by
\be \label{eq:Tands_charged}
 T = \frac{d(1-q)}{4\pi r_h} \ , \qquad s = \frac{1}{4G_5} \frac{\ell^{d-1}}{r_h^{d-1}} \ ,
\ee
while the physical charge, conjugate to the chemical potential $\mu$, is found by computing the derivative of the on-shell action with respect to $\mu$:
\be \label{eq:Qexp}
Q = \frac{\partial S_\mathrm{on-shell}}{\partial \mu}\bigg |_{T\ \mathrm{fixed}} = \frac{1}{16\pi G_5} \frac{(d-2)\ell^{d-3}}{r_h^{d-2}} \mu = \frac{1}{16\pi G_5} \frac{\sqrt{2d(d-1)}\ell^{d-2}}{r_h^{d-1}}\sqrt{q} \ .
\ee

\subsection{Analytic entanglement entropy for charged black holes}\label{sec:charged_anal}

It is immediate that the large $D$ approximation for the entanglement entropy of strips in charged black hole backgrounds is obtained from the above analysis by changing the expression for the blackening factor. In particular, the NH blackening factor of~\eqref{eq:fNH} becomes in the charged case
\be \label{eq:fRNNH}
 f_q(w) = (1-w)(1-qw) + \morder{\frac{1}{d}}
\ee
if $w$ is held fixed in the limit of large $d$.
Note that the charge dependence mostly modifies the geometry near the horizon. The leading order NB geometry in~\eqref{eq:NBmetric} is independent of charge and remains unchanged. However, the corrections due to the blackening factor do depend on the charge. Therefore the expansion~\eqref{eq:BHexactser} changes. Since the charged blackening factor has more complicated series expansion than the neutral one~\cite{Garcia-Garcia:2015emb}, we only write down the first two terms (corresponding to setting $k_c=1$ in the series~\eqref{eq:NBneutral} of the neutral case):
\begin{align} \label{eq:RNser}
\begin{aligned}
 A_\nb(r_*,r_c,q) &= \frac{V_2\ell^{d-1}}{r_c^{d-2}}\Bigg[-\frac{2}{d-2}  \, _2F_1\left(\frac{1}{2},-\frac{d-2}{2 (d-1)};\frac{d}{2 (d-1)};\left(\frac{r_c}{r_*}\right)^{2 d-2}\right)+&\\ 
&+ \frac{1}{2}   \left(1+\frac{d q}{d-2}\right) \left(\frac{r_c}{r_h}\right)^d\, _2F_1\left(\frac{1}{2},\frac{1}{d-1};\frac{d}{d-1};\left(\frac{r_c}{r_*}\right)^{2 d-2}\right)\Bigg]
 \ ,  & \\ 
 L_\nb(r_*,r_c,q) &= \frac{2r_c^d}{dr_*^{d-1}}\,\Bigg[ _2F_1\left(\frac{1}{2},\frac{d}{2 (d-1)}; \frac{3d-2}{2(d-1)};\left(\frac{r_c}{r_*}\right)^{2 d-2}\right) +&\\
 &+ \frac{1}{4}\left(1+\frac{d q}{d-2}\right)  \left(\frac{r_c}{r_h}\right)^d\, _2F_1\left(\frac{1}{2},\frac{d}{d-1};\frac{2d-1}{d-1};\left(\frac{r_c}{r_*}\right)^{2 d-2}\right)\Bigg] \ . &
\end{aligned}
\end{align}

\begin{figure}[hbt!]
\centering
\includegraphics[width=0.403\textwidth]{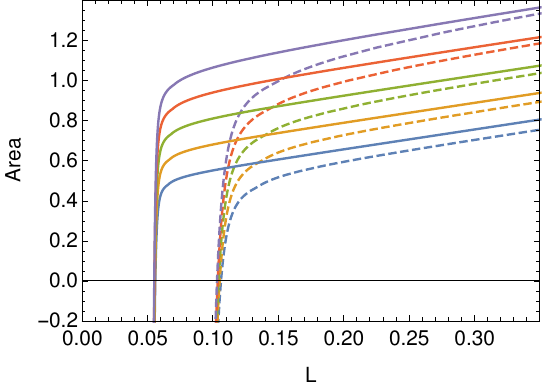}\hspace{4mm}%
\includegraphics[width=0.569\textwidth]{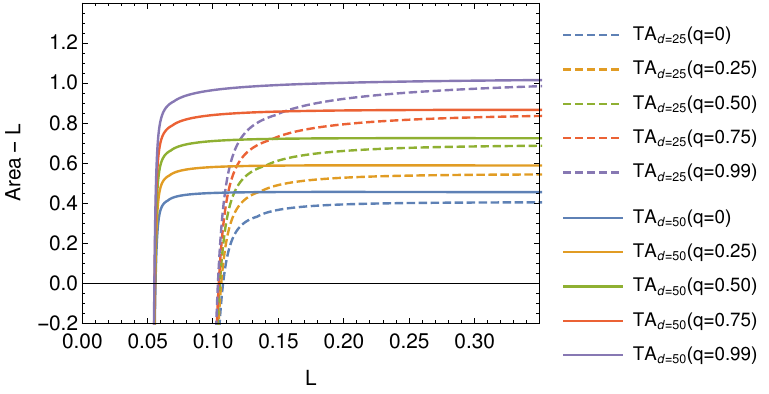}

\caption{Entanglement entropy (left) and the length-subtracted entanglement entropy (right) as a function of the length for different values of the charge $q=0$, $0.25$, $0.5$, $0.75$, and $0.99$,  in the analytic approximation. The values of the dimension are $d=25$ (dashed curves) and $d=50$ (solid curves).}
\label{fig:charge dependence}
\end{figure}

The near horizon expressions are modified more drastically due to the presence of the charge. Inserting the leading order blackening factor~\eqref{eq:fRNNH} in~\eqref{eq:NHintegrals}, the integrals evaluate to
\begin{align} \label{eq:NHcharged}
\begin{aligned}
 A_\nh(w_*,w_c,q) &= \frac{2V_2 r_h}{d w_* \sqrt{(1+w_*)(1-qw_*)}} \left(\frac{\ell}{r_h}\right)^{d-1} \times & \\
 &\ \ \times \Bigg[\frac{1}{w_c}\sqrt{\frac{(1-qw_*)(1-qw_c)(1+w_*) (w_*-w_c) (w_*+w_c)}{1-w_c}}+&\\
&\ \ \ \ \ \ +\left(1+w_*^2\right) F\left(\varphi|m\right)-(1+w_*)(1-qw_*) E\left(\varphi|m\right)+&\\
 &\ \ \ \ \ \  +(1+q)w_*(1-w_*)  \Pi \left(\frac{2}{1+w_*};\varphi\Big|m\right)\Bigg] \ , 
 & \\ 
 L_\nh(w_*,w_c,q) &= \frac{4 r_h}{d\sqrt{(1+w_*)(1-qw_*)}} F\left(\varphi|m\right)=\frac{4r_h\, \text{dn}^{-1}\left(\sqrt{\frac{(1-w_*) (1-q w_c)}{(1-w_c) (1-q w_*)}}\big|m\right)}{d \sqrt{(1+w_*)(1-qw_*)}} \ , &
\end{aligned}
\end{align}
where the angle is unchanged with respect to the neutral case, but the parameter is modified:
\be \label{eq:varphidef}
 \varphi = \sin ^{-1}\left(\sqrt{\frac{\left(1+w_*\right) (w_*-w_c)}{2w_*(1-w_c)}}\right) \ , \qquad m = \frac{2(1-q) w_*}{(1+w_*)(1-qw_*)}\ .
\ee
The final matched result is then obtained as above:
\begin{align}\label{eq:chargedfinal}
\begin{aligned}
 \mathrm{Area}(r_*,q) &= \left\{
 \begin{array}{lc}
    A_\nb(r_*,r_*,q)\ ,  & \qquad (r_* \le r_c)  \\
    A_\nb(r_*,r_c,q) +A_\nh(r_*^d,r_c^d,q)\ , &  \qquad (r_* > r_c) 
 \end{array}
 \right. & \\
 L(r_*,q) &= \left\{
 \begin{array}{lc}
    L_\nb(r_*,r_*,q)\ ,  & \qquad (r_* \le r_c)  \\
    L_\nb(r_*,r_c,q) +L_\nh(r_*^d,r_c^d,q)\ , &  \qquad (r_* > r_c) 
 \end{array}
 \right. & 
\end{aligned}
\end{align}
with $r_c/r_h = 1 -\frac{1}{2d}(\log d -1)$.

We analyze the charge dependence of the result~\eqref{eq:chargedfinal} in Fig.~\ref{fig:charge dependence}. It turns out to be mild, with the main effect being a slight overall increase in the regulated area. Note that we chose as the charge of the most extremal black hole to be $q=0.99$ instead of $q=1$. This is because our expressions become badly defined for $q=1$. This case of exactly extremal black holes will be analyzed separately below.
We also show the difference between the area and the length (in appropriate units) in the right panel if the figure. This is related to the subleading term of the entanglement entropy when expanded in $1/L$, which we will discuss in Sec.~\ref{sec:expansion}.

\section{Extremal black holes and quantum criticality}\label{sec:extremal}

We then discuss our result for extremal black holes in the bulk geometry, obtained as the limit $q \to 1$ from the expressions in the previous section. Specifically, we may consider the five-dimensional metric in~\eqref{eq:metric5RN}, where the blackening factor is now
\be
 f_\mathrm{ext}(r) = 1 -\left(1+\frac{d}{d-2}\right)\left(\frac{r}{r_h}\right)^d + \frac{d}{d-2}\left(\frac{r}{r_h}\right)^{2d-2} \ .
\ee
The extremality of the black hole means implies that the blackening factor has a double root at the horizon,
\be
 f_\mathrm{ext}(r) = \frac{d(d-1)}{r_h^2}\left(r-r_h\right)^2 + \morder{(r-r_h)^3} \ ,
\ee
so that in particular the temperature is zero, as also seen from~\eqref{eq:Tands_charged}.
Inserting this expression in the expressions~\eqref{eq:metricD} and~\eqref{eq:metric5} for the metric, we see that the asymptotic IR geometry is AdS$_2\times \mathbb{R}^3$ (or AdS$_2\times \mathbb{R}^{d-1}$ for the $D$-dimensional metric). The presence of the AdS$_2$ factor indicates the appearance of a ``quantum critical'' region in the zero temperature limit~\cite{Liu:2009dm,Faulkner:2009wj,Iqbal:2011in,Alho:2013hsa}. The  AdS factor in the geometry is interpreted to signal the presence of a one dimensional IR CFT~\cite{Faulkner:2009wj,Edalati:2010hk}. 

\subsection{Analytic entanglement entropy for extremal black holes}

In principle, the results for extremal black holes are obtained from those given in Sec.~\ref{sec:charged_anal} by taking $q \to 1$. For the NB expressions, this is indeed straightforward. We obtain, using~\eqref{eq:RNser},
\begin{align} \label{eq:extser}
\begin{aligned}
 A_\nb^{(\mathrm{ext})}(r_*,r_c) &= \frac{V_2\ell^{d-1}}{r_c^{d-2}}\Bigg[-\frac{2}{d-2}  \, _2F_1\left(\frac{1}{2},-\frac{d-2}{2 (d-1)};\frac{d}{2 (d-1)};\left(\frac{r_c}{r_*}\right)^{2 d-2}\right)+&\\ 
&+ \frac{1}{2}   \left(1+\frac{d}{d-2}\right) \left(\frac{r_c}{r_h}\right)^d\, _2F_1\left(\frac{1}{2},\frac{1}{d-1};\frac{d}{d-1};\left(\frac{r_c}{r_*}\right)^{2 d-2}\right)\Bigg]
 \ ,  & \\ 
 L_\nb^{(\mathrm{ext})}(r_*,r_c) &= \frac{2r_c^d}{dr_*^{d-1}}\,\Bigg[ _2F_1\left(\frac{1}{2},\frac{d}{2 (d-1)}; \frac{3d-2}{2(d-1)};\left(\frac{r_c}{r_*}\right)^{2 d-2}\right) +&\\
 &+ \frac{1}{4}\left(1+\frac{d}{d-2}\right)  \left(\frac{r_c}{r_h}\right)^d\, _2F_1\left(\frac{1}{2},\frac{d}{d-1};\frac{2d-1}{d-1};\left(\frac{r_c}{r_*}\right)^{2 d-2}\right)\Bigg] \ , &
\end{aligned}
\end{align}
where we again only write down  the leading corrections due to the blackening factor. However, the NH results~\eqref{eq:NHcharged} become singular as $q \to 1$. Therefore, one should take a step back and look at the integrals~\eqref{eq:NHintegrals}. Actually, the blackening factor in the extremal case takes a simple form,
\be
 f_\mathrm{ext}(w) = (1-w)^2 + \morder{\frac{1}{d}},
\ee
so that we insert $\sqrt{f(w)} = 1-w$ in the integrals, eliminating one appearance of square roots. After this, the integrals turn out to be doable in terms of elementary functions, rather than elliptic functions:
\begin{align} \label{eq:NHext}
\begin{aligned}
 A_\nh^{(\mathrm{ext})}(w_*,w_c) &= \frac{2 V_2 r_h}{d} \frac{\ell^{d-1}}{r_h^{d-1}}\left(\frac{\sqrt{w_*^2-w_c^2}}{w_c w_*}+\frac{w_* \cos ^{-1}\left(\frac{w_c-w_*^2}{w_*(1-w_c)}\right)}{\sqrt{1-w_*^2}}+\cosh ^{-1}\left(\frac{w_*}{w_c}\right)\right)\ ,&\\
 L_\nh^{(\mathrm{ext})}(w_*,w_c) &= \frac{4 r_h\sin ^{-1}\left(\sqrt{\frac{(1+w_*) (w_*-w_c)}{2w_*(1-w_c) }}\right)}{d\sqrt{1-w_*^2}} = \frac{4 r_h}{d\sqrt{1-w_*^2}}\varphi \ , &
\end{aligned}
\end{align}
with the angle $\varphi$ defined as above in~\eqref{eq:varphidef}.
That is, the final results is given again as in~\eqref{eq:neutralfinal} or in~\eqref{eq:chargedfinal} but the expressions are a lot simpler, and no special functions are needed.

\begin{figure}[hbt!]
\centering
\includegraphics[width=0.48\textwidth]{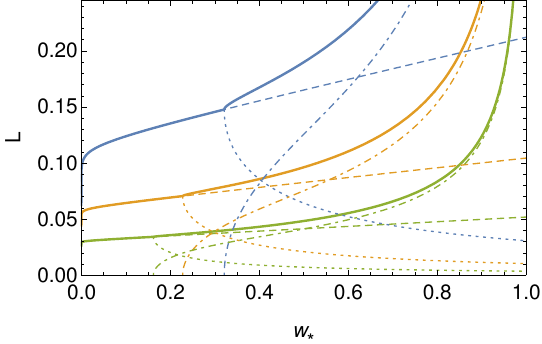}\hspace{4mm}%
\includegraphics[width=0.46\textwidth]{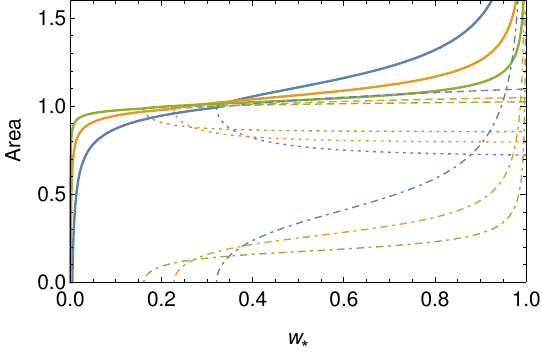}
\vspace{3mm}

\includegraphics[width=0.8\textwidth]{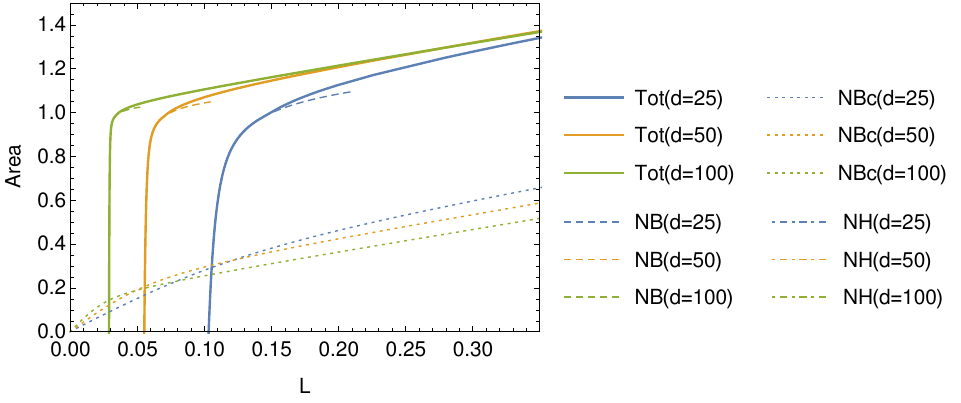}
\caption{Breakdown of the analytical approximation of the entanglement entropy to near-boundary and near-horizon contributions for extremal black holes. Notation as in Fig.~\ref{fig:breakdown_neutral}. Top left: Length of the strip as a function of the turning point $w_*$. Top right: Regulated area of the embedding as a function of $w_*$. Bottom: Area as a function of the length.}
\label{fig:breakdown_ext}
\end{figure}

\subsection{Results}

We show the results for the entanglement entropy in extremal black holes in Fig.~\ref{fig:breakdown_ext}. Similarly as for neutral black holes in Fig.~\ref{fig:breakdown_neutral}, the figure includes the final results as well as the breakdown into various contributions. Notation is the same as in Fig.~\ref{fig:breakdown_neutral}. When the area is viewed as a function of the strip length (bottom plot), there differences with respect to the neutral case are small. This is in agreement with Fig.~\ref{fig:charge dependence}, where $q=0$ is the neutral case and $q \to 1$ is the extremal limit. However, as a function of $w_*$ (panels in the top row) the difference with respect to neutral black holes is larger. It is most pronounced when the embeddings approach the horizon, $w_* \to 1$. This reflects the different near-horizon behavior of the geometry in the extremal black hole case.

\section{Soliton backgrounds}\label{sec:soliton}

Finally, we consider entanglement entropy in the so-called soliton backgrounds at large $d$. This means considering backgrounds in the $D$-dimensional Einstein gravity~\eqref{eq:highDS} having the form
\be \label{eq:metricDsol}
 ds_D^2 = \frac{\ell^2}{r^2}\left(\frac{dr^2}{f(r)} - dt^2 + dx_1^2 +dx_2^2 +f(r)dx_3^2 +d\mathbf{y}^2\right) \ ,
\ee
where the third spatial coordinate $x_3$ is compactified in order to make the geometry regular~\cite{Witten:1998zw}, the coordinates $y_i$ cover the $D-5=d-4$ dimensional flat transverse manifold, and $f(r)$ is given in~\eqref{eq:fexp}. That is, the role of time and $x_3$ are exchanged with respect to the black hole geometry~\cite{Aharony:2006da}. The dimensionally reduced geometry is also similar to the dimensionally reduced black hole geometry~\eqref{eq:metric5}:
\be \label{eq:metric5sol}
 ds_5^2 = \left(\frac{\ell}{r}\right)^{\frac{2}{3}(d-1)}\left(\frac{dr^2}{f(r)}- dt^2  + dx_1^2 +dx_2^2 +f(r)dx_3^2\right) \ .
\ee
However, the expressions for the holographic entanglement entropy are modified because in the case of the soliton, the bulk embedding wraps $x_3$ while in the case of the black hole it does not wrap $t$. Importantly, this means that unlike in the case of black holes, there are two competing extremal surfaces. First, there is a ``connected'' surface, which is qualitatively similar to the extremal surface in the black hole case. But now there is also a ``disconnected'' surface, which consists of two straight pieces of surfaces hanging from the end points of the strip, and ending at the point where $f(r)$ vanishes (which we shall still call $r_h$ even if there is no horizon). These pieces are in principle connected through another piece at fixed $r=r_h$, but this piece has zero area and does not contribute to the entropy. We will here focus on the connected surface, as the results for the disconnected surfaces are trivial. 
The computation is mostly similar to the black hole case, and therefore we suppress much of the details below. 

\subsection{Analytic entanglement entropy for soliton backgrounds}

Following the steps outlined above, we find for the NB expressions
\begin{align} \label{eq:solser}
\begin{aligned}
 A_\nb(r_*,r_c) &= \frac{V_2\ell^{d-1}}{r_c^{d-2}}\Bigg[-\frac{2}{d-2}  \, _2F_1\left(\frac{1}{2},-\frac{d-2}{2 (d-1)};\frac{d}{2 (d-1)};\left(\frac{r_c}{r_*}\right)^{2 d-2}\right)+&\\ 
&+\frac{1}{2 d}\left(\frac{r_c}{r_*}\right)^{2d-2}\left(\frac{r_c}{r_h}\right)^d \, _2F_1\left(\frac{3}{2},\frac{d}{d-1};\frac{2d-1}{d-1};\left(\frac{r_c}{r_*}\right)^{2 d-2}\right)-&\\
 &-\frac{1}{d}\left(\frac{r_c}{r_*}\right)^{d-2}\left(\frac{r_c}{r_h}\right)^d \, _2F_1\left(\frac{3}{2},\frac{d}{2 (d-1)};\frac{3d-2}{2(d-1)};\left(\frac{r_c}{r_*}\right)^{2 d-2}\right)\Bigg]
 \ ,  & \\ 
 L_\nb(r_*,r_c) &= \frac{2r_c^d}{dr_*^{d-1}}\,\Bigg[ _2F_1\left(\frac{1}{2},\frac{d}{2 (d-1)}; \frac{3d-2}{2(d-1)};\left(\frac{r_c}{r_*}\right)^{2 d-2}\right) +&\\
 & +\frac{1}{2} \left(\frac{r_c}{r_h}\right)^d \, _2F_1\left(\frac{3}{2},\frac{d}{d-1};\frac{2d-1}{d-1};\left(\frac{r_c}{r_*}\right)^{2 d-2}\right)-&\\
 &-\frac{1}{2} \left(\frac{r_*}{r_h}\right)^d \, _2F_1\left(\frac{3}{2},\frac{d}{2 (d-1)};\frac{3d-2}{2(d-1)};\left(\frac{r_c}{r_*}\right)^{2 d-2}\right)-&\\
 &
 -\frac{d }{8 d-4}\left(\frac{r_c}{r_*}\right)^{3d-2}\!\left(\frac{r_*}{r_h}\right)^d\! \, _2F_1\left(\frac{3}{2},\frac{2d-1}{d-1};\frac{3d-2}{d-1};\left(\frac{r_c}{r_*}\right)^{2 d-2}\right)\Bigg]  &
\end{aligned}
\end{align}
where we included only the leading correction.

The NH expressions become 
\begin{align} \label{eq:NHsol}
\begin{aligned}
 A_\nh(w_*,w_c) &= \frac{2V_2 r_h}{d w_* } \left(\frac{\ell}{r_h}\right)^{d-1} \Bigg[\frac{1}{w_c}\sqrt{(1-w_c) (w_*+w_c-w_*w_c) (w_*-w_c)}+ & \\
 & +E\left(\left.\varphi\right|m\right)-(1-w_*) F\left(\left.\varphi\right|m\right)\Bigg] \, , 
 & \\ 
 L_\nh(w_*,w_c) &= 
 \frac{2r_h \sqrt{1-w_*} }{d}\text{dn}^{-1}\left(\sqrt{\frac{1-w_*}{1-w_c}}\Big|m\right)
 \ , &
\end{aligned}
\end{align}
where now
\be
 \varphi = \sin ^{-1}\left(\sqrt{\frac{w_c+w_*-w_*w_c}{1-(1-w_*)^2}}\right) \ ,  \qquad  m = 1-(1-w_*)^2 \ .
\ee

As it turns out, there is a significant difference in the matching relation between $r_*$ and $w_*$ with respect to the black hole geometry. The counterpart of the derivative expressions in~\eqref{eq:derexps} now reads
\be \label{eq:derexpssol}
 \left.\frac{dx_1}{dr}\right|_\mathrm{NB} = \frac{1}{\sqrt{\frac{r_*^{2d-2}}{r^{2d-2}}-1}} \ , \qquad \left.\frac{dx_1}{dr}\right|_\mathrm{NH} = d \frac{r^{d-1}}{r_h^d}\frac{r_h}{d}\frac{1}{w \sqrt{1-w}\sqrt{\frac{w_*^{2}}{w^{2}}\frac{1-w}{1-w_*}-1}} \ ,
\ee
i.e., the NB expression is unchanged but the NH expression contains extra factors arising from $f(r)$. Applying the NB and NH approximations simultaneously, i.e., approximating $w^{1/d}\approx 1$ and neglecting power corrections on $w$, we obtain
\be
 \left.\frac{dx_1}{dr}\right|_\mathrm{NB} \approx \frac{1}{\sqrt{\frac{r_*^{2d-2}}{r_h^{2d-2}w^2}-1}}\ , \qquad \left.\frac{dx_1}{dr}\right|_\mathrm{NH} \approx \frac{1}{\sqrt{\frac{w_*^{2}}{w^{2}(1-w_*)}-1}} \ .
\ee
Therefore matching now leads to the relation 
\be \label{eq:matchingsol}
 \frac{w_*}{\sqrt{1-w_*}} = \left(\frac{r_*}{r_h}\right)^d \ ,
\ee
which differs from the naive expectation from the coordinate definition~\eqref{eq:wdef} and the corresponding result~\eqref{eq:srel} for the black hole geometry. However, this relation has a minor drawback: since it does not agree with the coordinate definition~\eqref{eq:wdef} in general, in particular it differs from it at the gluing point $r_*=r_c$. That is, one cannot have both $w_* \to w_c$ and $r_* \to r_c$ as one approaches the gluing point. This leads to the results for the length and the area of the embedding being discontinuous at this point as $r_*$ varies. The discontinuity is suppressed at large $d$, but in order to make the results cleaner we introduce a modified matching formula which removes it. To do this, instead of dropping the factor of $1-w$ inside the latter square root factor in the NH formula in~\eqref{eq:derexpssol}, we approximate it by the value $1-w_c$ at the gluing point. Here $w_c \sim 1/d$, so the correction is suppressed at large $d$. We obtain
\be \label{eq:matchingsoli}
 \frac{w_*\sqrt{1-w_c}}{\sqrt{1-w_*}} = \left(\frac{r_*}{r_h}\right)^d \ .
\ee

The final matched result can now be written as
\begin{align}\label{eq:solfinal}
\begin{aligned}
 \mathrm{Area}(r_*) &= \left\{
 \begin{array}{lc}
    A_\nb(r_*,r_*)\ ,  & \qquad (r_* \le r_c)  \\
    A_\nb(r_*,r_c) +A_\nh(w_*(r_*,r_c),r_c^d)\ , &  \qquad (r_* > r_c) 
 \end{array}
 \right. & \\
 L(r_*) &= \left\{
 \begin{array}{lc}
    L_\nb(r_*,r_*)\ ,  & \qquad (r_* \le r_c)  \\
    L_\nb(r_*,r_c) +L_\nh(w_*(r_*,r_c),r_c^d)\ , &  \qquad (r_* > r_c) 
 \end{array}
 \right. & 
\end{aligned}
\end{align}
where $w_*(r_*,r_c)$ is solved from~\eqref{eq:matchingsoli},
\be
 w_*(r_*,r_c) = \frac{1}{2\left(1-\left(\frac{r_c}{r_h}\right)^d\right)}\left(\frac{r_*}{r_h}\right)^{2d}\left[\sqrt{1+4\left(1-\left(\frac{r_c}{r_h}\right)^d\right)\left(\frac{r_h}{r_*}\right)^{2d}}-1\right] \ .
\ee
The matching location can again to be chosen to be $r_c/r_h = 1 - (\log d -1)/(2d)$.

\begin{figure}[hbt!]
\centering
\includegraphics[width=0.8\textwidth]{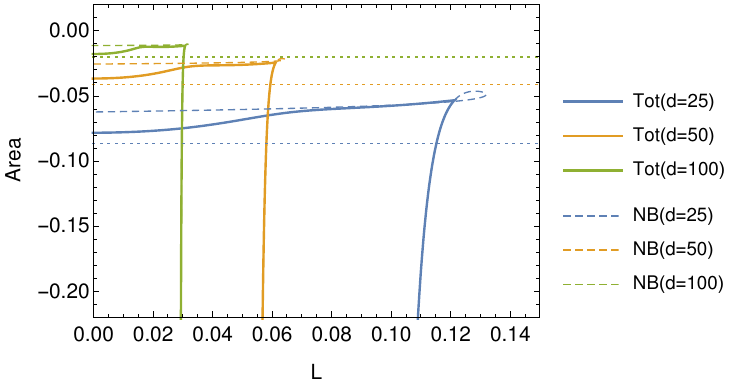}
\caption{Area as a function of the length in the analytical approximation for soliton geometries.}
\label{fig:LvsA_soliton}
\end{figure}

\subsection{Results}

We show the regulated area as a function of the length for soliton geometries with three different choices of $d$, as estimated by the formulas~\eqref{eq:solfinal}, in Fig.~\ref{fig:LvsA_soliton}. Following notation in Figs.~\ref{fig:breakdown_neutral} and~\ref{fig:breakdown_ext}, the thick solid curves give the full result for the connected surface and dashed curves show the result using only the NB expression. The dotted horizontal lines are the results for the disconnected surface, given by
\begin{equation}
    \textrm{Area} = -\frac{2V_2\ell^{d-1}}{(d-2)r_h^{d-2}} \ .
\end{equation}
There are two branches\footnote{{The terminology here should be understood such that the stable solution is the global minimum in the Ryu-Takayanagi description, whereas the unstable solution is only a local minimum and should be discarded when evaluating the final result.}} of the connected solution, ``stable'' and ``unstable''. The stable solutions in Fig.~\ref{fig:LvsA_soliton} are shown by the nearly vertical curves which enter from the bottom of the plot.  
For small $L$, this stable solution coincides with the NB contribution (dashed curves) for each $d$, as 
expected in this region. 
The unstable solutions are shown by the nearly horizontal curves which lie above the disconnected results  (the disconnected results are marked by the dotted lines). 
The total results and the  
NB contribution mismatch for the unstable branch because the turning point enters the NH region.

Only the stable branch of the connected solution and the disconnected solution contribute to the final result for the entanglement entropy.  
The details in the unstable branch of the connected solution, including the behavior near the turning point and the wiggles in the nearly horizontal branch in the figure, are irrelevant for the final result.

\begin{figure}[hbt!]
\centering
\includegraphics[width=0.6\textwidth]{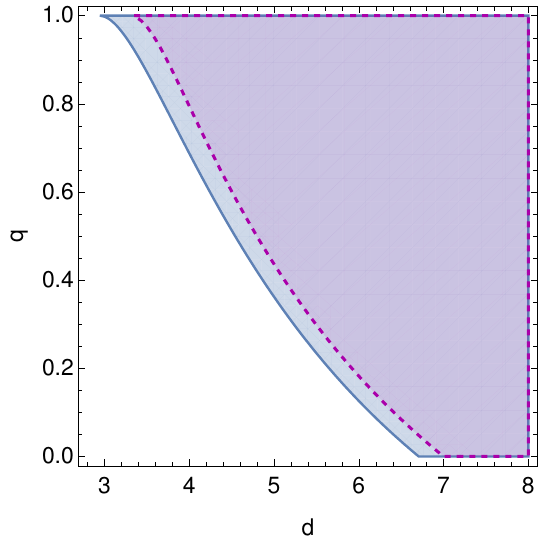}

\caption{The region on the ($d$,$q$)-plane where area theorem is violated. The blue solid curve and shaded area so the exact result obtained numerically, whereas the dashed violet and shade show the area according to our large-$d$ approximation.}
\label{fig:violation}
\end{figure}

\section{Large width expansion of the entanglement entropy} \label{sec:expansion} 

The entanglement entropy for strips (say in five dimensions) may be expanded at large strip length $L$ as
\begin{equation}\label{eq:SELexp}
    S_E(L) = s V_2 L + S_0 + \morder{\frac{1}{L}}  \ ,
\end{equation}
where $V_2$ is the volume in the spatial directions transverse to the strip, and $s$ is the thermal entropy density. Indeed, it is a standard relation for holographic entanglement that the leading ``volume'' term can be identified as the thermal entropy in the region of entanglement, and this can be show to hold also hold in field theory using a lattice formulation~\cite{Jokela:2023rba}. The subleading ``area'' term $S_0$ is known to have interesting properties: In general, one can write it as $S_0 = \alpha\ \text{vol}\partial A$, where $\alpha$ is monotonic under Lorentz-covariant flows so that $\alpha_\mathrm{IR} < \alpha_\mathrm{UV}$~\cite{Casini:2012ei,Casini:2016udt}. 
This property is known as the area theorem. 
%%%%%%%%%%%%%%%%%%%%%%%%%%%%%%%%%%%%%%%%%%
The area theorem is inspired from the c-theorem which  tells us the behaviour of the entanglement entropy under an RG flow. As the central charge is linked with the entanglement entropy, we expect that for a fixed length,   $c_\mathrm{IR} < c_\mathrm{UV}$ which in turn means monotonic decrease in the entanglement entropy as we move from the UV fixed point to the IR fixed point. But, if $\alpha_\mathrm{UV} - \alpha_\mathrm{IR}$ is negative then this area theorem gets violated as the entanglement entropy no longer decreases under the RG flow.
%%%%%%%%%%%%%%%%%%%%%%%%%%%%%%%%%%%%%%%%%%%%
Indeed, it is known that the area theorem does not hold for non-Lorentz-covariant flows and has shown to be ``violated'' in many holographic setups implementing such flows~\cite{Gushterov:2017vnr,Jeong:2022zea}. In particular, it is known to be violated by (field theories dual to) black holes at large $d$. In this case, the violation of the theorem is present if the coefficient $S_0$ is positive in the black hole background~\cite{Giataganas:2021jbj}.

Following this recipe, the region where the $S_0>0$ for charged black holes can be computed numerically (by evaluating the difference $S_E(L)-s V_2 L$ using the expressions~\eqref{eq:BHexactreg} at large $L$). We consider here the general case of charged black holes: neutral or extremal cases are obtained simply by setting the charge to zero or to one, respectively, in the blackening factor~\eqref{eq:fexpq}. We show the result as a function of the dimension $d$ and the charge $q$ 
 in blue in Fig.~\ref{fig:violation}. 
 We compare the result to the estimate obtained from the analytic formulas~\eqref{eq:chargedfinal}, shown as the violet shaded region. Despite the curve being at rather low $d$, the analytic estimate is rather good. Note also that $S_0$ is positive at $d=4$ (i.e., four-dimensional CFTs) for highly charge black holes, and even in the case $d=3$ if one picks black holes very near extremality.

\subsection{Analytic result for strips at large $D$}

It is also possible to further analyze subleading term $S_0$ analytically in the limit of large $D$~\cite{Giataganas:2021jbj}, without resorting to the approximation of~\eqref{eq:chargedfinal}. Again we consider black holes with a generic charge $q$ as the results for the neutral or extremal black holes are obtained simply by taking $q \to 0$ or $q \to 1$. In order to obtain the analytic expression, we first consider the difference
\be \label{eq:DeltaA}
 \Delta\mathrm{Area}(r_*,q)\equiv\mathrm{Area}(r_*,q)-V_2 \left(\frac{\ell}{r_h}\right)^{d-1}L(r_*,q) -\mathrm{Area}_\mathrm{str}(q) 
\ee
in the limit $r_* \to r_h$, where $\mathrm{Area}_\mathrm{str}(q)$ is the area of a straight $d-1$ dimensional surface between the boundary at the horizon. Since divergences cancel in the difference, we can consider either regulated or unregulated expressions for the areas. The unregulated area for the straight surface is given by 
\be
 \mathrm{Area}_\mathrm{str}(q) = 2 V_2 \int_{\epsilon}^{r_h}dr\, g_{xx}(r)g_{rr}(r)^{1/2}= 2 V_2\ell^{d-1}\,\int_{\epsilon}^{r_h}dr\,\frac{1}{r^{d-1}\sqrt{f_q(r)}} \ .
\ee
Note that the difference~\eqref{eq:DeltaA} measures the error of approximating the minimal area by a rectangular surface (since the second term is the area of the horizon piece of the rectangle). 
Without any approximations, we obtain the integral
\begin{align}
\begin{aligned}
\Delta\mathrm{Area}(r_*=r_h,q) &= 2 V_2\left(\frac{\ell}{r_h}\right)^{d-1}\,\int_0^{r_h}dr\,\frac{r_h^{d-1}}{r^{d-1}\sqrt{f_q(r)}}  \left(\sqrt{1-\frac{r^{2d-2}}{r_h^{2d-2}}}-1\right) & \\
&=2 V_2\left(\frac{\ell}{r_h}\right)^{d-1}\frac{r_h}{d} \int_0^1 dw\, \frac{w^{2/d}}{w^2\sqrt{f_q(w)}}\left(\sqrt{1-w^{2-2/d}}-1\right) \ .
\end{aligned}
\end{align}
We note that, unlike for example the regulated area integral in~\eqref{eq:BHexactreg}, this integral is finite even if we set $d \to \infty$. Therefore we have that\footnote{In the counting of factors of $1/d$ in this section, the ``trivial'' factor of $\ell^d/r_h^d$ which appears in all expressions is left unexpanded.}
\be \label{eq:DeltaAfinal}
\Delta\mathrm{Area}(r_*=r_h,q) = \morder{\frac{1}{d}} \ .
\ee

Let us then study the regulated expression for the straight area,
\begin{align} \label{eq:stAreareg}
\begin{aligned}
 \mathrm{Area}_\mathrm{str}(q) &= 2 V_2\ell^{d-1}\,\int_{0}^{r_h}dr\,\frac{1}{r^{d-1}}\left(\frac{1}{\sqrt{f_q(r)}}-1\right) - \frac{2 V_2\ell^{d-1}}{(d-2)r_h^{d-2}} & \\
 &= 2 V_2\left(\frac{\ell}{r_h}\right)^{d-1}\frac{r_h}{d}\,\int_{0}^{1}dw\,\frac{w^{2/d}}{w^2}\left(\frac{1}{\sqrt{1-\left(1+\frac{dq}{d-2}\right)w+\frac{dq}{d-2}w^{2-2/d}}}-1\right) - &\\ 
 &
 \ \ \ - \frac{2 V_2\ell^{d-1}}{(d-2)r_h^{d-2}}\ .&
\end{aligned}
\end{align}
Note that naively, the result appears to be in $\mathcal{O}(1/d)$ as both the terms include explicit factors of $1/d$ or $1/(d-2)$. However, the $w$-integral becomes logarithmically divergent if one takes $d \to \infty$ in the integrand, which prevents simple power counting. The divergence can be isolated by adding and subtracting a term in the integral: 
\begin{align} \label{eq:stAreareg2}
\begin{aligned}
 \mathrm{Area}_\mathrm{str}(q) 
 &= 2 V_2\left(\frac{\ell}{r_h}\right)^{d-1}\frac{r_h}{d}\,\int_{0}^{1}dw\,\frac{w^{2/d}}{w^2}\times &\\
 &\qquad \times\left(\frac{1}{\sqrt{1-\left(1+\frac{dq}{d-2}\right)w+\frac{dq}{d-2}w^{2-2/d}}}-\frac{1}{2}\left(1+\frac{dq}{d-2}\right)w-1\right) + &\\ 
 &
 \ \ + V_2\left(\frac{\ell}{r_h}\right)^{d-1}\frac{r_h}{d}\left(1+\frac{dq}{d-2}\right)\int_0^1 dw\,\frac{w^{2/d}}{w} - \frac{2 V_2\ell^{d-1}}{(d-2)r_h^{d-2}}\ .&
\end{aligned}
\end{align}
Now the first integral is well-behaved as $d \to \infty$, so its contribution is indeed $\mathcal{O}(1/d)$, but the divergence does appear in the second integral in the limit $d \to \infty$. However, this integral can be immediately computed, with the perhaps surprising result that it is $\mathcal{O}(d^0)$. The divergence is only regulated due to the numerator $w^{2/d}$ at exponentially small $w \sim e^{-d}$, which gives rise to the enhancement by a factor of $d$. 
The explicit result is given by  
\be \label{eq:Astrfinal}
 \mathrm{Area}_\mathrm{str}(q) =   \frac{V_2\ell^{d-1}}{2r_h^{d-2}}(1+q) + \morder{\frac{1}{d}}\ .
\ee

Combining the results~\eqref{eq:DeltaAfinal} and~\eqref{eq:Astrfinal}, we find for the coefficient in the expansion for the area (which only differs from $S_0$ by $1/4G_5$) 
\be \label{eq:strip_S0_final}
 A_0 \equiv \lim_{r_*\uparrow r_h }\left[\mathrm{Area}(r_*,q)-V_2 \left(\frac{\ell}{r_h}\right)^{d-1}L(r_*,q)\right] =  \frac{V_2\ell^{d-1}}{2r_h^{d-2}}(1+q) + \morder{\frac{1}{d}} \ .
\ee
This expression is in agreement with the numerical results in Fig.~\ref{fig:charge dependence} (right): $\text{Area}-L$ at large $L$ is given by to $(1+q)/2$ in naturally chosen units.

Finally, the leading corrections to the integrals can also be found analytically. A straightforward computation gives 
\begin{align} \label{eq:strip_S0_subleadings}
\begin{aligned}
 A_0 &=  \frac{V_2\ell^{d-1}}{2r_h^{d-2}}(1+q) + & \\ & \quad +
 \frac{V_2\ell^{d-1}}{dr_h^{d-2}}\left[1-2 \sqrt{2}\sqrt{1-q}+(1+q) \log \frac{ 12-4q-8\sqrt{2} \sqrt{1- q}}{(1+q)^2}\right]+\morder{\frac{1}{d^2}} \ .&
\end{aligned}
\end{align}

\subsection{Generalization to other entanglement regions}

The formula~\eqref{eq:strip_S0_final} gives an exact result for the value of the expansion coefficient $S_0$ at large $d$, but the result perhaps does not appear to be very illuminating. However,  studying the derivation more closely, we can make the following interesting observations:
\begin{enumerate}
    \item The result arose from the divergence of the second integral in~\eqref{eq:stAreareg2}. This divergence in turn can be traced back to the leading correction in the blackening factor in~\eqref{eq:stAreareg}. The coefficient in this vacuum expectation value (VEV) term has an independent physical interpretation as the difference between the internal and free energy densities, $\eps-f = sT+\mu Q$.
    \item While the derivation was carried out in the case of the strip, it is clear that the result generalizes to other regions, in the limit where their size is taken to be large. This is because the final result~\eqref{eq:strip_S0_final} arose from the near boundary divergence in the expression for the area of the minimal surface, and near the boundary, the $r$-dependence of the minimal surface was so weak, that it could be replaced by the straight surface. The same is expected to hold for other regions than the strip.
\end{enumerate}
As a simple check, note that using the expressions in~\eqref{eq:Tands_charged} and~\eqref{eq:Qexp} we find that
\be
 sT+\mu Q =\frac{d}{16\pi G_5} \frac{\ell^{d-1}}{r_h^{d-2}}\left(1+\frac{dq}{d-2}\right)  = \frac{d}{16\pi G_5} \frac{\ell^{d-1}}{r_h^{d-2}}(1+q) + \morder{d^{0}} 
\ee
which is indeed proportional to~\eqref{eq:strip_S0_final}. Therefore, the entanglement entropy in the case of the strip becomes\footnote{Since here we took first the limit of wide strip and then the limit of large $d$, a natural question to ask is whether the order of limit matters. As it turns out, it does not, but it is simpler to take $L \to \infty$ first.}
\be
 S_E(L) = s V_2 L  + \frac{4\pi V_2}{d}\left( sT+\mu Q\right) + \morder{\frac{1}{d},\frac{1}{L}} \ .
\ee
That is, at large $d$, apart from the well-known leading order result, also the subleading term in the $1/L$ expansion could be written in terms of thermodynamic variables.

Without attempting a precise proof, we therefore expect that the entanglement entropy for sufficiently ``large'' regions $A$ with smooth boundaries is given by
\be \label{eq:SElargeA}
 S_E(A) \approx \mathrm{Vol}(A) s + \mathrm{Vol}(\partial A) \frac{2\pi}{d} \left( sT+\mu Q\right)
\ee
with corrections suppressed by $1/d$ and the characteristic length scale of $A$. To make this latter correction precise we can choose first a fixed smooth region $\widehat A$, and define a uniformly scaled region (with $\lambda>0$)
\be \label{eq:Areascaling}
 A(\lambda) = \left\{x \in \mathbb{R}^3\,|\, x/\lambda \in \widehat A\right\}
\ee
and consider the limit $\lambda \to \infty$. That is, using this definition,~\eqref{eq:SElargeA} holds up to corrections suppressed by $1/\lambda$. As far as we can see, $A_0$ does not need to be connected or bounded. As for the smoothness requirement, it is clear that the structure of the boundary must be limited by some minimal length scale for the scaling in~\eqref{eq:Areascaling} to lead to a region where all distances are large as $\lambda \to \infty$. In particular, self-similar structures are not allowed. But it is apparently enough for the boundary to be piecewise smooth: defects in the surface are one-dimensional objects, and therefore will lead to contributions that scale as $\mathrm{Vol}(\partial A)/\lambda$ and are subleading in~\eqref{eq:SElargeA} as $\lambda \to \infty$.

Some additional remarks are in order. First, note that the thermodynamical interpretation of the VEV coefficient of the blackening function implicitly requires that our expressions are valid close to the boundary. This means that for a more general near-critical setup, where the large-$d$ approach only describes the geometry near the IR, additional care is needed to correctly interpret the result. However, since $T$ and $s$ are defined at the horizon, the interpretation of the result in terms of $sT$ is still valid, but the term $\mu Q$ is less obvious. Second, if we stick to the conformal case, we can use the equation of state for a CFT, $\epsilon = -(d-1)f$ to write $\eps -f = -f\, d$. Then, in terms of the free energy density, the formula takes an even simpler form:
\be \label{eq:SEalt}
 S_E(A) \approx \mathrm{Vol}(A) s - \mathrm{Vol}(\partial A) 2\pi f \ .
\ee
Note that $f=\morder{d^0}$ in our counting, while $\epsilon = \morder{d}$. Moreover, the difference $\epsilon-f$ is  free of UV divergences, whereas the free energy needs to be renormalized, so this form for the expansion assumes the standard regularization of $f$ (i.e., subtraction of the vacuum energy).

We also note that the large-$d$ result is already partially visible in the NB expansions we discussed above. That is, the term $k=1$ in the expansion for the area in~\eqref{eq:BHexactser} enhanced by a factor of $d$ with respect to all other terms as $d \to \infty$ (see the factor $d(k-1)+2$ in the denominator). Consequently, the result for ($A_0$ and) $S_0$ arises from this term in the expansion~\cite{Giataganas:2021jbj}.  This is in agreement with the above observation that $A_0$ is proportional to the VEV term in the blackening factor. The importance of the $k=1$ term at $d \to \infty$ is also the main reason why we included it in the NB part of the matched expressions (e.g.,~\eqref{eq:neutralfinal}) rather than only using the pure AdS $k=0$ term.

Moreover, the above discussion is restricted to areas in 3+1 dimensional field theory. A natural question is whether the formula~\eqref{eq:SElargeA}, or some modified version of the formula, also holds for generic regions $A$ the $d$-dimensional CFT, before carrying out the dimensional reduction. This is nontrivial because curvature effect in higher dimension may affect the NB analysis leading to this formula.

\section{Conclusions}\label{sec:conclusions}

In this article, we applied standard large $D$ methods to analyze holographic entanglement entropy. We focused on AdS$_D$ black hole geometries in the Poincar\'e patch, dual to finite temperature CFTs in $D-1$ dimensions, and to the simplest case of entanglement regions, i.e., strips. We demonstrated that the method indeed applies in this case: one can obtain precise analytic approximations to the entanglement entropy via applying the Ryu-Takayanagi prescription separately in the near-boundary and near-horizon regions of the geometry, and by matching the results in the middle. We were able to obtain analytic results for neutral, charged, and extremal black holes. 

Our analysis complement earlier results~\cite{Garcia-Garcia:2015emb,Giataganas:2021jbj} for entanglement entropy in black hole geometries which were based on (arbitrary order) series expansions of the blackening factor. These earlier results essentially give the near-boundary terms in our approach. In the case of strips, such near-boundary expansions can describe (as we show in Fig.~\ref{fig:near_brdy_exp}) the nontrivial part of the functional dependence of the entropy on the width of the strip, $S_E(L)$. Therefore, one might think that adding the near-horizon result does not really add much to the approach. However note that by adding it, one obtains an expression which convergences to the exact function $S_E(L)$ uniformly (rather than pointwise) in $L$ as $D \to \infty$. But, perhaps more importantly, we argued in Sec.~\ref{sec:setup} and in the Appendix that our method works also in the case of general nearly-critical geometries which only agree with (dimensionally reduced) AdS black holes in the near-horizon regions when the black holes are small enough. In such cases, it is the near-horizon part of the result for the entanglement entropy which is unmodified with respect to the pure AdS black hole case, whereas the near-boundary part is modified. Even if (as we show in Appendix) the effect of the modification on the final result is small in the large $D$ limit, the modification complicates applying direct near-boundary series expansion of the blackening factors.

We also analyzed entanglement entropy in the limit of wide strips. Interestingly, the subleading correction to the entropy could be computed analytically at large $D$. We pointed out that this term arises from a logarithmically enhanced term near the boundary. Therefore, we argued that the result is universal and applies to all sufficiently smooth regions in the limit where their size is taken large.

There are various future directions to explore. We focused here on the simplest region, the strip, but one could apply the method to other regions as well. However, this will be more challenging because the equation of motion for the embedding is easily integrable only in the case of strips. Nevertheless it should be possible at least to check explicitly that our expansion in system size~\eqref{eq:SElargeA} holds also in the case of other regions.

Since our result here (see Eqs.~\eqref{eq:SElargeA}) and~\eqref{eq:SEalt}) was expressed in terms of thermodynamic quantities, even in the presence of charge in the geometry, its variation can be readily studied by using standard thermodynamic formulas. One should be able to understand how this result relates to the proposed (generalized) first law of the entanglement entropy (see, e.g.,~\cite{DiNunno:2021eyf} and references therein).

The large $D$ expansion is in principle expandable to higher order. As we pointed out, in the case of AdS spaces, all order corrections the NB result are already known analytically. The bottleneck is therefore the computation of the NH results. However, it seems that it will be challenging to obtain analytic closed form expressions for the corrections in this region. Therefore, in order to extend the method to higher orders, one needs to write the result in terms of integrals, for example.

Another direction is to apply the method to other observables. A simple example, considered in~\cite{Giataganas:2021jbj}, is the Wilson loop. The Wilson loop is computed in holography by evaluating the action of a string anchored to the loop at the boundary, which should lead to similar expressions as the entanglement entropy. A different possibility would be to apply our method to study timelike entanglement entropy~\cite{Doi:2023zaf}. It may also be interesting to study other observables related to quantum information such as complexity.

One can also consider other geometries than the plain AdS black holes. The natural extension would be to work in the lower dimensional setup, and consider other dilaton potentials with near-critical exponential IR asymptotics than the purely exponential choice. This typically means that the NB embeddings cannot be solved analytically. The method as described here only works as an approximation for small black holes: the NH results (which are new results in this article) hold still exactly, but the expressions for the NB embeddings only work in the region which is deep in the IR and far from the horizon of the small BH. Nevertheless, the method should give a reasonable approximation for the full result. Note however that there is a specific setup where full analytic control can be maintained~\cite{Betzios:2017dol,Betzios:2018kwn,Gursoy:2021vpu}. Namely, one can glue a section of AdS$_5$ directly to the nearly critical BH geometry (i.e., the geometry obtained by dimensionally reducing the AdS$_D$ black hole), so long as the gluing point is far enough from the BH horizon. In this case, the full NB embedding can still be computed analytically even though the geometry is rather complex. A related open question is the fate of the results of Sec.~\ref{sec:expansion}, e.g., Eq.~\eqref{eq:SElargeA} in this kind of more general geometries. It should be checked whether this result generalizes in some form to these geometries that are nearly-critical in the IR but not exactly given by dimensionally reducing higher dimensional AdS black holes.

Naturally, it would also be extremely interesting, albeit probably challenging, to check whether~\eqref{eq:SElargeA} also holds on the CFT side. To start with, one could try to check the result for a strip in the limit of large width or high temperature. This could provide a nice additional check of the Ryu-Takayanagi formula for holographic entanglement entropy.

\acknowledgments

We thank for discussions M.~Afrasiar, J.~K.~Basak, D.~Giataganas, N.~Jokela, B.-H. Lee, M.~Lippert, J.~Pedraza, R.~Meyer, H.~Ruotsalainen, F.~Subandi, R.~Suzuki, and M.~Watanabe.  P.~J. and M.~J. have been supported by an appointment to the JRG Program at the APCTP through the Science and Technology Promotion Fund and Lottery Fund of the Korean Government. P.~J. and M.~J. have also been supported by the Korean Local Governments --- Gyeong\-sang\-buk-do Province and Pohang City --- and by the National Research Foundation of Korea (NRF) funded by the Korean government (MSIT) (grant number 2021R1A2C1010834).

\appendix

\section{On the generalization to UV-complete nearly critical geometries}\label{app:critical}

In this Appendix we argue that our method applies also 
nearly critical to $3+1$ dimensional setups with AdS$_5$ asymptotics near the boundary.
For concreteness, we consider black hole geometries and the entanglement entropy for a strip. Similar arguments hold for other shapes.

We assume that the potential is asymptotically AdS$_5$ near the boundary, which is obtained for example from an extremum of the potential,
\be \label{eq:potUVas}
 V(\phi) = -\frac{12}{\ell^2} + m^2\left(\phi-\phi_*\right)^2 + \cdots
\ee
where the dots denote terms of higher order in $\phi-\phi_*$.
We consider potentials which agree asymptotically with the expression~\eqref{eq:Vreduced} obtained by a dimensional reduction of $d+1$ dimensional Einstein gravity,
\be \label{eq:potIRas}
 V(\phi) = -  \frac{d(d-1)}{\tilde \ell^2} \exp\left[\frac{4}{3}\sqrt{\frac{d-4}{d-1}}\phi\right]\,\left(1+\cdots \right) 
\ee
where the dots denote terms suppressed at large $\phi$.
Here the IR radius $\tilde \ell$ is expected to scale as $\tilde \ell \sim d$ at large $d$ so that the coefficient of the potential remains finite (see~\cite{Betzios:2018kwn,Gursoy:2021vpu}).
Recall that for a generic five-dimensional metric~\eqref{eq:5dgenmetric}, the expressions for the strip length and area are given in~\eqref{eq:stripintegrals}. Next we shall analyse these expressions for the potentials with asymptotics~\eqref{eq:potUVas} and~\eqref{eq:potIRas}.

We discuss here the geometry using the conformal coordinates,
\be
 \label{eq:confcoords}
 ds_5^2 = e^{2A(r)}\left(\frac{dr^2}{f(r)}-f(r) dt^2 + d\mathbf{x}^2\right) \ .
 \ee
Near the boundary, the geometry is asymptotically AdS$_5$,
\be \label{eq:nearbdrymetric}
 ds_5^2 = \frac{\ell^2}{r^2}\left(dr^2 + \eta_{\mu\nu}dx^\mu dx^\nu\right)\left[1+\morder{\left(\frac{r}{r_\mathrm{UV}}\right)^2}\right] \ ,
\ee
where $r_\mathrm{UV}$ sets the scale where deviation from the AdS metric becomes large. Going deeper in the bulk, one encounters the horizon of the black hole at some point. If the black hole is small enough, the geometry will be determined by the asymptotics in~\eqref{eq:potIRas}, and is that of a $D$ dimensional back hole reduced to five dimensions,~\eqref{eq:metric5}. We stress that for our approximation to make sense, we need to require that the black hole horizon lies deep enough in the geometry so that using the asymptotic form~\eqref{eq:potIRas} is enough to capture the geometry.

There is however a complication when combining the expressions~\eqref{eq:nearbdrymetric} and~\eqref{eq:metric5}. That is, the coordinate $r$ is not necessarily the same, but may be shifted between the two expressions, and the shift may be large. We fix the freedom in shifting such that $r$ vanishes at the boundary, so that~\eqref{eq:nearbdrymetric} holds without change. Then, requiring that the UV and the IR asymptotics join smoothly, we find that~\eqref{eq:metric5} should actually be replaced by  
\be \label{eq:dsIRform}
 ds_5^2 = \left(\frac{\tilde \ell}{r+\delta r +\tilde \ell}\right)^{\frac{2}{3}(d-1)}\left(\frac{dr^2}{f(r)}-f(r) dt^2 + d\mathbf{x}^2\right)\left[1+\morder{\left(\frac{r_\mathrm{UV}}{r}\right)^{\#}}\right]
\ee
where the subleading term of the geometry depends on the subleading terms of the potential and might not be a power law but, for example, logarithmically suppressed in $1/r$. Here we only need to now that the corrections grow large for $r \sim r_\mathrm{UV}$. Note that the holographic coordinate has been shifted by $\tilde \ell +\delta r$, where $\delta r \sim d^0$ at large $d$, whereas $\tilde \ell \sim d$. The value of the $\mathcal{O}(1)$ correction does not affect our analysis so we will set $\delta r =0$ in the following. Then (the leading term of) the blackening factor reads
\be
 f(r) = 1 -\left(\frac{r+\tilde\ell}{r_h +\tilde \ell}\right)^d
\ee
where $r=r_h$ is the location of the black hole horizon. Requiring the horizon to lie in the IR regime means that $r_h \gg r_\mathrm{UV}$.

We then check whether the results for entanglement entropy from the IR geometry~\eqref{eq:dsIRform} are useful to describe the full result. In the conformal coordinates, the integrals in~\eqref{eq:stripintegrals} may be written as
\begin{align} \label{eq:stripintegralsconfA}
 \textrm{Area}(r_*) &= 2 V_2\,\int_{\epsilon}^{r_*}dr\,\frac{e^{3A(r)}}{\sqrt{f(r)}\sqrt{1-\exp[6A(r_*)-6A(r)]}} \ , & \\ L(r_*) &= 2\, \int_{0}^{r_*}dr\,\frac{e^{3A(r_*)}}{e^{3A(r)}\sqrt{f(r)}\sqrt{1-\exp[6A(r_*)-6A(r)]}} \ , &
\label{eq:stripintegralsconfL} 
\end{align}
where we introduced a cutoff regularization for the area.
Note that if the turning point $r_*$ in these formulas is close enough to the horizon $r_h$, both the integrals are dominated by the near-horizon contribution: the latter square root factor in the denominators gives rise to a pole at $r=r_*$ as $r_* \to r_h$. Therefore it is enough to check the result when $r_*$ is not close to $r_h$. 

Note that when $r_*$ is not close to $r_h$, the square root factors in the integrand for the length in~\eqref{eq:stripintegralsconfL} give $\morder{1}$ contributions, so the dominant terms are the $e^A$ factors. Therefore the integrand behaves as $\propto e^{-3A}$.  As $e^{A}$ decreases fast with increasing $r$ both near the boundary and deeper in the IR, it is immediate that the length integral is dominated by the IR contributions. The integral for the area is however more tricky: the integrand behaves as $e^{3A}$, so that it is obviously dominated by the near boundary contribution, which naively appears to prevent any useful estimates of the area using the IR geometry only. However, the near boundary contribution turns out to be a trivial constant. That is, the dependence on the horizon $r_h$ and turning point $r_*$ appears\footnote{To be precise, the function  $A(r)$ also includes corrections depending on  $r_h$ and therefore on the temperature, but these corrections have the same behavior as the explicit $f(r)$ dependence.} through the square root factors in~\eqref{eq:stripintegralsconfA}. For $r_* \gg r_\mathrm{UV}$, the contribution due to the latter square root factor is
\be
 \sim \int_\epsilon^{r_*}dr\ \frac{e^{6A(r_*)}}{e^{3A(r)}}  
\ee
as found by expanding the square root factor as a series. This is (similarly to the expression for $L$) a strongly IR dominated correction. The temperature corrections are however more delicate: expanding the blackening factor in~\eqref{eq:stripintegralsconfA} we find 
\be \label{eq:ftcorr}
 \sim \int^{r_*}dr\ e^{3A(r)}\left(\frac{r+\tilde\ell}{r_h +\tilde \ell}\right)^d \sim  \int^{r_*}dr\  \frac{\tilde \ell^{d-1}(r+\tilde \ell)}{\left(r_h +\tilde \ell\right)^d}
\ee
where we inserted the IR form of the geometry to obtain the final expression.
This integral is also IR dominated so long as $r_* \gg r_\mathrm{UV}$, but the increase of the integrand is weaker than in the other cases, so its essential that $r_* \gg r_\mathrm{UV}$.

We note that these results also hold for the vacuum geometry for which $f(r)=1$, with the understanding that the correction of Eq.~\eqref{eq:ftcorr} is simply absent, since none of the steps above make use of any specific form of $f(r)$.

In summary, we find that the IR contribution is a good estimate to the full entanglement entropy, up to a trivial constant, whenever the minimal surface extends significantly to the IR region. That is,
\begin{itemize}
    \item For the $T=0$ vacuum geometry, the entanglement entropy for the strip as a function of the length is well produced when the strip is long enough to probe the IR geometry, which means $L \gg r_\mathrm{UV}$.
    \item The temperature dependence is captured for black holes with $r_h \gg r_\mathrm{UV}$ for long enough strips ($L\gg r_\mathrm{UV}$).
\end{itemize}
Finally we note that the steps in the derivation do not require using the large $d$ limit for the IR geometry, so the conclusions are valid even at finite $d$. However, we only find analytic results for the entanglement entropy for the IR geometries in the large $d$ limit.

\bibliographystyle{JHEP}
\bibliography{references.bib}

\end{document}